\pdfoutput=1
\documentclass{article}

\usepackage{arxiv}

\usepackage[utf8]{inputenc} 
\usepackage[T1]{fontenc}    
\usepackage{hyperref}       
\usepackage{url}            
\usepackage{booktabs}       
\usepackage{amsfonts}       
\usepackage{nicefrac}       
\usepackage{microtype}      
\usepackage{lipsum}		
\usepackage{graphicx}
\usepackage{natbib}
\usepackage{doi}

\title{Diffuse ultrasound computed tomography for medical imaging}

\author{ \hspace{1mm}Ines Elisa Ulrich\footnote{Corresponding author: \texttt{ines.ulrich@erdw.ethz.ch}} \\
	ETH Zürich\\
	Department of Earth Sciences\\
	Institute of Geophysics\\
	Zürich CH-8092, Switzerland\\
	\texttt{ines.ulrich@erdw.ethz.ch} \\
	\And
	Christian Boehm \\
	ETH Zürich\\
	Department of Earth Sciences\\
	Institute of Geophysics\\
	Zürich CH-8092, Switzerland\\
	\texttt{christian.boehm@erdw.ethz.ch} \\
	\And
	Andrea Zunino \\
	ETH Zürich\\
	Department of Earth Sciences\\
	Institute of Geophysics\\
	Zürich CH-8092, Switzerland\\
	\texttt{andrea.zunino@erdw.ethz.ch} \\
	\And
	Cyrill Bösch \\
	ETH Zürich\\
	Department of Earth Sciences\\
	Institute of Geophysics\\
	Zürich CH-8092, Switzerland\\
	\texttt{cyrill.boesch@erdw.ethz.ch}\\
	\And
	Andreas Fichtner \\
	ETH Zürich\\
	Department of Earth Sciences\\
	Institute of Geophysics\\
	Zürich CH-8092, Switzerland\\
	\texttt{andreas.fichtner@erdw.ethz.ch}\\
}

\renewcommand{\shorttitle}{Diffuse ultrasound computed tomography for medical imaging}

\hypersetup{
pdftitle={A template for the arxiv style},
pdfsubject={q-bio.NC, q-bio.QM},
pdfauthor={David S.~Hippocampus, Elias D.~Striatum},
pdfkeywords={First keyword, Second keyword, More},
}

\begin{document}
\maketitle

\begin{abstract}
	An alternative approach to ultrasound computed tomography (USCT) for medical imaging is proposed, with the intent to (i) shorten acquisition time for devices with a large number of emitters, (ii) eliminate the calibration step, and (iii) suppress instrument noise. Inspired by seismic ambient field interferometry, the method rests on the active excitation of diffuse ultrasonic wavefields and the extraction of deterministic travel time information by inter-station correlation. To reduce stochastic errors and accelerate convergence, ensemble interferograms are obtained by phase-weighted stacking of observed and computed correlograms, generated with identical realizations of random sources. Mimicking a breast imaging setup, the accuracy of the travel time measurements as a function of the number of emitters and random realizations can be assessed both analytically and with spectral-element simulations for realistic breast phantoms. The results warrant tomographic reconstructions with straight- or bent-ray approaches, where the effect of inherent stochastic fluctuations can be made significantly smaller than the effect of subjective choices on regularisation. This work constitutes a first conceptual study and a necessary prelude to future implementations.
	
	The following article has been submitted to Journal of the Acoustical society of America. After it is published, it will be found at \url{http://asa.scitation.org/journal/jas}.
\end{abstract}

\keywords{Inverse Theory \and Green's function retrieval \and diffuse wavefield correlation \and ultrasound tomography}

\section{Introduction}
Being cost-efficient and free of ionizing radiation, ultrasound computed tomography (USCT) is an attractive medical imaging modality that receives increasing attention. Early on, \citet{GreenleafJohnson1975}, \citet{GreenleafBahn1981} and \citet{Glover1977} demonstrated the potential of using transmission ultrasound to image breast tissue. However, partly due to limited computational power, the development of USCT for breast screening stagnated during the following decades. It only recently regained momentum with the development of new USCT devices with a large number of emitter-receiver pairs that provide dense angular coverage \citep{KIT_3Dscanner, QT_MalikTerryWiskin2018, SoftVue_Duric2013}. Correlating the direct arrivals of the pressure wavefield in a reference medium (typically water) and the tissue of interest, provides a set of travel time differences that may be used in a ray-based inversion to constrain sound speed variations. The reference recordings additionally serve as calibration dataset that suppresses the source imprint on the waveforms that would otherwise corrupt the travel time measurements. The combination of correlation-based travel time measurements and the ray-based tomography has the advantage of being robust and efficient, which is essential for the high frequencies of $1$ to several MHz that current USCT devices operate with. Several studies have shown that this setup may produce sound speed maps of human breast tissue that are qualitatively comparable to mammography or MRI scans \citep{SoftVue_MRIcomparison,KIT_MRIcomparison}.  

Despite the undeniable success and promise, USCT faces serveral challenges that may affect its integration into a highly optimized clinical routine. Since the acquisition time of a USCT dataset scales with the number of emitters, a dense emitter-receiver array that produces the desired high-resolution images may becomes impractical when it records longer than the typical time scales of patient movement \citep{RoyDuric}. The acquisition time may be shortened by reducing the number of repeated shots, however, at the expense of a decreased signal-to-noise ratio in the resulting stack. Last but not least, the calibration step requires the acquisition of an additional dataset, which complicates the processing, further extends the total acquisition time, and potentially acts as an additional source of errors.

With the primary goals of (i) shortening acquisition time, (ii) suppressing instrument noise, and (iii) eliminating calibration measurements, we investigate an alternative approach whereby travel time information for tomography is extracted from correlations of diffuse ultrasonic wavefields. This is inspired by the approximation of inter-receiver Green's functions by diffuse wavefield correlations that can be shown theoretically \citep[e.g.,][]{Claerbout,LobkinsWeaver2001,Wapenaar2003,WapenaarFokkema2006} and experimentally \citep[e.g.,][]{MalcolmScalesTiggelen2004}, and is widely used in seismology to image the Earth's crust and mantle on the basis of ambient seismic noise  \citep[e.g.,][]{ShapiroCampillo2005,Sabra_2005,Stehly_2009,Saygin_2012,Nakata_2019b}. Since ambient ultrasonic wavefields are not available, we propose to generate a quasi-random wavefield actively which can then be used for the estimation of inter-receiver travel times from cross correlations of individual recordings. Here, the term random wavefield specifically referres to the fact that a diffuse wavefield is characterized by random uncorrelated modal amplitudes with equal variances (in literature, this characteristic is sometimes termed wavefield equipartitioning).

In the following, we derive the relevant equations for Green's function retrieval by diffuse wavefield correlation and present its translation from the large-scale passive seismic setup to the small-scale active medical setup. To reduce stochastic errors, we suggest to generate observed and computed wavefields with identical realizations of random sources, and to accelerate convergence of the ensemble correlations by phase-weighted stacking.

As the simplest proof of concept, we first study a homogeneous medium where a random wavefield can be computed analytically. This serves to estimate the accuracy of travel time estimates. Subsequently, we consider numerically computed random wavefields propagating through a heterogeneous medium, which we then reconstruct tomographically using both straight- and bent-ray algorithms. We conclude with a discussion of the advantages and disadvantages of the method, which should outline the niche within which it may be beneficial.

In the current absence of USCT devices that could implement random wavefield interferometry, this work is a first conceptual study and a necessary prelude to future practical considerations.

This work makes an effort to provide reproducible science as introduced by the Stanford Exploration Project. Figures labelled [R] are reproducible using codes and input files provided in \footnote{\url{https://mybinder.org/v2/gl/swp_ethz\%2Fpublic\%2Frandom-field-interferometry/master}}.

\section{Theory}\label{sec:theory}

To set the stage, we consider a setup where the domain of interest $\Omega$ consists of a water bath with the immersed human breast enclosed by a transducer array, which holds the ultrasonic transducers and delineates the boundary of the domain. For a circular frequency $\omega$, the acoustic pressure $p(\mathbf{x},\omega)$ at position $\mathbf{x}$ is related to the speed of sound $c(\mathbf{x})$ and mass density $\rho(\mathbf{x})$ of the medium by the acoustic wave equation 
\begin{equation}
 \frac{\omega^2}{\rho(\mathbf{x}) c^2(\mathbf{x})}p(\mathbf{x},\omega)+\nabla\cdot\bigg(\frac{1}{\rho(\mathbf{x})}\nabla p(\mathbf{x},\omega)\bigg)=-\frac{1}{\rho(\mathbf{x})}f(\mathbf{x},\omega),
 \label{eq:wave_equation_frequency_domain}
\end{equation}
where the volumetric force density gradient $f(\mathbf{x},\omega)$ of the emitting transducers acts as external source. Neumann, Dirichlet or absorbing boundary conditions may be enforced along different parts of the domain boundary, depending on the specifics of a particular setup. We further assume that the medium is at rest prior to the action of the sources.

\subsection{Diffuse wavefield interferometry}\label{SS:Interferometry}
To derive the relevant equations for Green's function retrieval by diffuse wavefield correlation, we borrow an argument from the normal-mode theory by \citet{Weaver_2004}. A collection of alternative derivations may be found in \citet{Fichtner_2019b}. As a starting point, we expand the $j^\text{th}$ realization of a random pressure wavefield $p_j(\mathbf{x},\omega)$ into the normal modes $\phi_n(\mathbf{x})$ of the acoustic wave operator as
\begin{equation}
    p_j(\mathbf{x},\omega)=\sum_n a_{j,n}(\omega)\phi_n(\mathbf{x}),
    \label{eq:modal_expansion_p}
\end{equation}
where $a_{j,n}(\omega)$ are the frequency-dependent expansion coefficients or modal amplitudes computed as (see Appendix)
\begin{equation}
    a_m(\omega)=-\frac{1}{(\omega^2-\omega_m^2)}  \int_{\Omega}\frac{1}{\rho(\mathbf{x})}\phi_m^{\ast}(\mathbf{x})f(\mathbf{x},\omega)d\mathbf{x}\,.
    \label{eq:expansion_coeffs_arbirtary_source}
\end{equation}
The normal modes form a complete basis, satisfy the boundary conditions and are orthonormal under a weighted inner product, which is detailed in the Appendix. Using eq.\,(\ref{eq:modal_expansion_p}), the cross-correlation interferogram $C_j(\mathbf{x}_A,\mathbf{x}_B,\omega)$ between the recordings of the $j^\text{th}$ realization of the diffuse wavefield at positions $\mathbf{x}_A$ and $\mathbf{x}_B$ can be written as a double sum over normal modes,
\begin{align}
    \begin{split}
    C_j(\mathbf{x}_A,\mathbf{x}_B,\omega)&=p_j^{\ast}(\mathbf{x}_A,\omega)p_j(\mathbf{x}_B,\omega)=\bigg[\sum_n a_{j,n}(\omega) \phi_n(\mathbf{x}_A)\bigg]^{\ast}\bigg[\sum_m a_{j,m}(\omega)\phi_m(\mathbf{x}_B)\bigg]\\
    &=\sum_n\sum_m a_{j,n}^{\ast}(\omega) a_{j,m}(\omega)\phi_n^{\ast}(\mathbf{x}_A)\phi_m(\mathbf{x}_B).
    \label{eq:C_j_xA_xB}
    \end{split}
\end{align}
Taking the arithmetic mean over $N$ realizations yields the ensemble correlation 
\begin{align}
    \begin{split}
    C(\mathbf{x}_A,\mathbf{x}_B,\omega)&=\frac{1}{N}\sum_{j=1}^N C_j(\mathbf{x}_A,\mathbf{x}_B,\omega) = \frac{1}{N}\sum_{j=1}^N\sum_n\sum_m a_{j,n}^{\ast}(\omega)a_{j,m}(\omega)\phi_n^{\ast}(\mathbf{x}_A)\phi_m(\mathbf{x}_B)\\
    &=\frac{1}{N}\sum_n\sum_m\phi_n^{\ast}(\mathbf{x}_A)\phi_m(\mathbf{x}_B)\sum_{j=1}^N a_{j,n}^{\ast}(\omega) a_{j,m}(\omega).
    \label{eq:ensemble_average_xcorr}
    \end{split}
\end{align}
With some approximation, the sum over random realizations $j$ in (\ref{eq:ensemble_average_xcorr}) can be eliminated under the assumption of equipartitioning, meaning that modal amplitudes are on average nearly uncorrelated in the sense of
\begin{equation}
    \frac{1}{N}\sum_{j=1}^N a_{j,n}^{\ast}(\omega)a_{j,m}(\omega)\approx \gamma_n(\omega)\delta_{mn}\,,
    \label{eq:equipartitioning}
\end{equation}
with a frequency-dependent average modal power spectrum $\gamma_n(\omega) = \frac{1}{N} \sum_{j=1}^N |a_{j,n}(\omega)|^2$. Substituting (\ref{eq:equipartitioning}) into (\ref{eq:ensemble_average_xcorr}) yields a simplified, approximate expression for the ensemble correlation,
\begin{equation}\label{eq:ensemble}
C(\mathbf{x}_A,\mathbf{x}_B,\omega) \approx \sum_n \gamma_n(\omega) \phi_n^{\ast}(\mathbf{x}_A)\phi_n(\mathbf{x}_B)\,.
\end{equation}
Eq.\,(\ref{eq:ensemble}) may be compared to the normal-mode representation of the Green's function between $\mathbf{x}_A$ and $\mathbf{x}_B$ \citep{Gilbert_normal_modes},
\begin{equation}
    G(\mathbf{x}_A,\mathbf{x}_B,\omega)=-\sum_n\frac{1}{(\omega^2-\omega_n^2)}\phi_n^{\ast}(\mathbf{x}_B)\phi_n(\mathbf{x}_A)\,,
    \label{eq:GF_normal_mode}
\end{equation}
which we derive, for completeness, in the Appendix. 

The comparison of (\ref{eq:ensemble}) and (\ref{eq:GF_normal_mode}) reveals the well-known proportionality of the ensemble correlation $C(\mathbf{x}_A,\mathbf{x}_B,\omega)$ to the Green's function $G(\mathbf{x}_A,\mathbf{x}_B,\omega)$ in the frequency domain. The real-valued proportionality factor depends on the modal power spectrum $\gamma(\omega)$, which is controlled by the frequency content and spatial distribution of the wavefield sources. In most cases, $\gamma_n(\omega) \neq (\omega^2 - \omega_n^2)^{-1}$, meaning that the ensemble correlation is not exactly equal to the Green's function, even in the hypothetical case of infinitely many realizations. This difference has implications for the measurement of travel time differences for tomographic reconstructions, which we will consider in more detail in section \ref{SS:measurement}.

Importantly, the phases of the active sources that generate the diffuse wavefield are absent from eq.\,(\ref{eq:ensemble}). Hence, the cross-correlation eliminates unknown time shifts of the source wavelets, which typically require calibration runs prior to the actual experiment. Furthermore, the accumulation of a sufficiently accurate ensemble correlation may require less acquisition time than the successive firing of all individual sources one by one.

In the following sections, we focus on the extraction of travel time information from ensemble correlations. For this, we employ the time-domain version of $C(\mathbf{x}_A,\mathbf{x}_B,\omega)$, denoted by $C(\mathbf{x}_A,\mathbf{x}_B,t)$.

\subsection{Practical calculation of correlation functions}\label{SS:practical_calculation_of_xcorr}

In seismology, where noise interferometry is used extensively, diffuse wavefields are generated by ambient sources, such as ocean waves and atmospheric turbulence \citep[e.g.][]{Ardhuin_2011,Ermert_2017,Gualtieri_2019,Ardhuin_2019,Nakata_2019b,Igel_2021}, or anthropogenic activity. Due to the absence of comparable passive sources in medical ultrasound, we propose to generate random wavefields actively. 

For this, all transducers act simultaneously, each transmitting a different random source time function for a time interval of length $T$. This produces the first random wavefield realization $p_1(\mathbf{x},t)$ and the first set of inter-receiver correlations $C_1(\mathbf{x}_A,\mathbf{x}_B,t)$. Repeating this process $N$ times with new realizations of source time functions, yields the random wavefield realizations $p_2(\mathbf{x},t), ..., p_N(\mathbf{x},t)$ and corresponding correlations $C_2(\mathbf{x}_A,\mathbf{x}_B,t), ..., C_N(\mathbf{x}_A,\mathbf{x}_B,t)$. The ensemble correlation $C(\mathbf{x}_A,\mathbf{x}_B,t)$ is then defined as the arithmetic mean
\begin{equation}\label{E:200}
C(\mathbf{x}_A,\mathbf{x}_B,t) = \frac{1}{N} \sum_{j=1}^N C_j(\mathbf{x}_A,\mathbf{x}_B,t)\,.
\end{equation}
As demonstrated in section \ref{SS:Interferometry}, we expect $C(\mathbf{x}_A,\mathbf{x}_B,t)$ to approximate the inter-receiver Green's function $G(\mathbf{x}_A,\mathbf{x}_B,t)$. The travel time between locations $\mathbf{x}_A$ and $\mathbf{x}_B$ may then - for instance - be estimated by picking the maximum amplitude peak in $C(\mathbf{x}_A,\mathbf{x}_B,t)$.

The number of realizations $N$ needed to obtain stable travel time estimates from (\ref{E:200}) may be impractically large. The convergence of $C(\mathbf{x}_A,\mathbf{x}_B,t)$ towards the inter-station Green's function may be accelerated by replacing the linear arithmetic mean in \eqref{E:200} with the phase-weighted average \citep{Schimmel_1997,Schimmel_2011}
\begin{equation}\label{E:201}
C_\text{pw}(\mathbf{x}_A,\mathbf{x}_B,t) = \frac{1}{N} w(\mathbf{x}_A,\mathbf{x}_B,t) \sum_{j=1}^N C_j(\mathbf{x}_A,\mathbf{x}_B,t)\,,
\end{equation}
where the phase weight $w(\mathbf{x}_A,\mathbf{x}_B,t)$ is defined as
\begin{equation}\label{E:202}
w(\mathbf{x}_A,\mathbf{x}_B,t) = \frac{1}{N} \left| \sum_{k=1}^N e^{i \varphi_k(\mathbf{x}_A,\mathbf{x}_B,t) }   \right|^\nu\,.
\end{equation}
The angle $\varphi_k(\mathbf{x}_A,\mathbf{x}_B,t)$ is the phase of the analytic signal $\tilde{C}_k=C_k + i \mathcal{H}(C_k)$, where $\mathcal{H}$ denotes the Hilbert transform. The exponent $\nu$ controls the extent to which coherent phases are emphasised, and is typically chosen between $1$ and $3$ in order to avoid excessive waveform distortions in this nonlinear averaging procedure \citep{Schimmel_1997,Schimmel_2011}. We adopt $\nu=2$ in all subsequent examples.

\begin{figure}
\includegraphics[width=\textwidth]{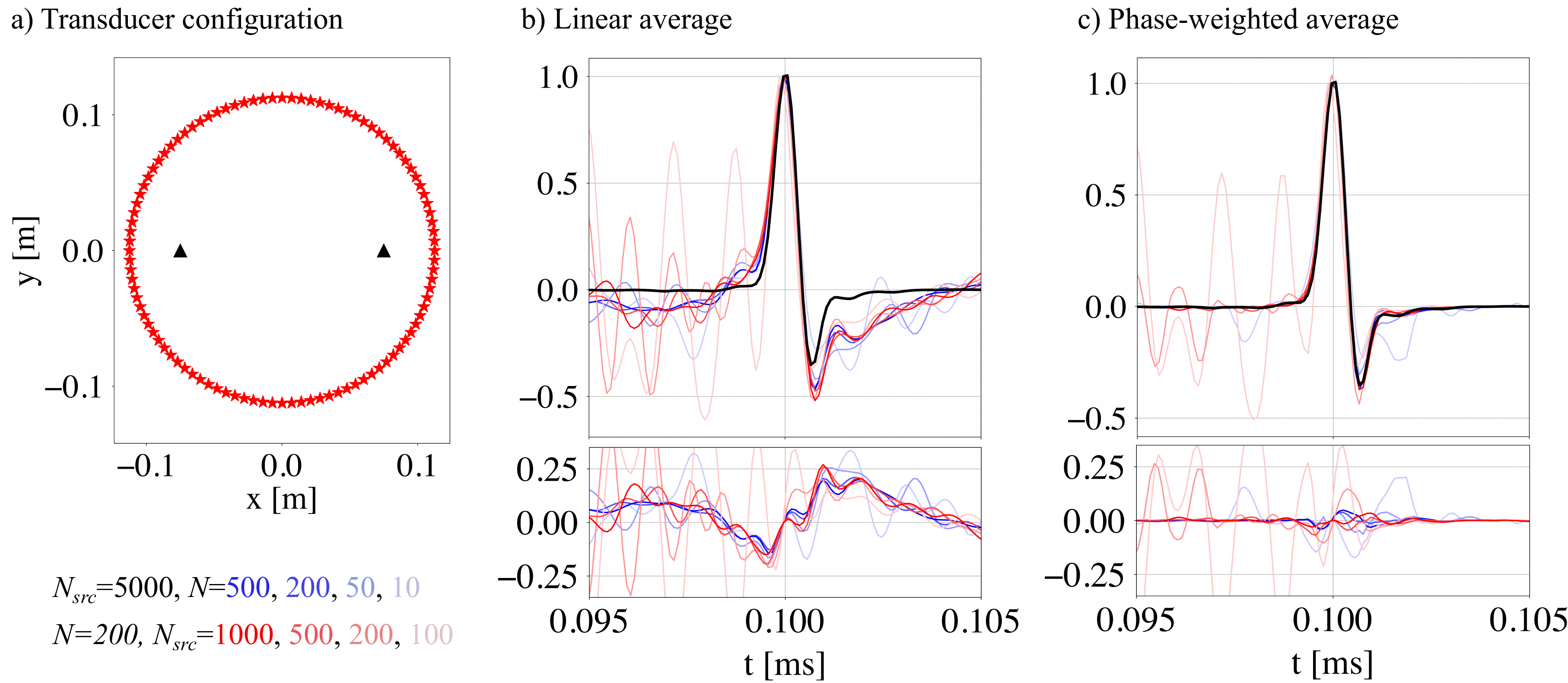}
\caption{Random wavefield correlations for a 2-D transducer ring. (a) Configuration of sources (red stars) and receivers (black triangles). The sources act simultaneously, each radiating a different random wavelet with a maximum frequency of $1$ MHz for the duration of $10$ ms, in order to produce one random wavefield realization. (b) Linear averages of the inter-receiver correlation for different numbers of realizations, $N$, and sources, $N_\text{src}$, along the ring. The analytical bandpass-filtered Green's function is plotted in black. The lower panel shows the error, that is, the difference between the analytical Green's function and the ensemble correlations, colour-coded according to the different $N$ and $N_\text{src}$. (c) Ensemble correlations similar to panel (b) but for the phase-weighted averaging, according to eq.\,(\ref{E:202}). The phase weight exponent was set to $\nu=2$.[R] (see introduction for the reproducibility indicator.)}
\label{F:summary_correlations}
\end{figure}

An illustrative example for the active generation of random wavefields and inter-receiver correlations is presented in Fig.\,\ref{F:summary_correlations}. The transducer configuration is two-dimensional, with sources arranged in a circle of $0.1$ m radius. For a duration of $10$ ms, the sources simultaneously radiate a random source time function, generated by a random time series, with a maximum frequency of $1$ MHz. The receiver positions $\mathbf{x}_A$ and $\mathbf{x}_B$ are at $r=15$ cm distance, so that the expected arrival time of the direct wave is $0.1$ ms for the homogeneous medium with sound speed $c=1500$ m$/$s. Fig.\,\ref{F:summary_correlations}b shows linear averages corresponding to eq.\,(\ref{E:200}) over varying numbers of random wavefield realizations, $N$, and sources along the 2-D ring, $N_\text{src}$. Comparing $C(\mathbf{x}_A,\mathbf{x}_B,t)$ to the analytical Green's function, given in 2-D by \citep{Igel}
\begin{equation}\label{eq:2DGreensfunction}
    G(r,t)=\frac{1}{2\pi c^2}U(t-\frac{|r|}{c})(t^2-\frac{r^2}{c^2})^{-\frac{1}{2}},
\end{equation}
with $U$ being the unit step function and $r=\sqrt{(x_A-x_B)^2+(y_A-y_B)^2}$, one notes - as expected - that the ensemble correlation approaches the analytical Green's function as $N$ and $N_\text{src}$ increase. The effect of the phase weight, shown in  \ref{F:summary_correlations}b, is to accelerate convergence, mostly by suppressing side-lobes around the main pulse.

\subsection{Travel time difference measurements}\label{SS:measurement}

Errors in the measurement of travel times may result from remaining random fluctuations for a finite number of realization, but also from differences between correlations and Green's functions, explained in section \ref{SS:Interferometry}. To reduce measurement errors, we propose to estimate travel time differences $\Delta t_{AB}$ directly instead of trying to measure absolute arrival times in wave pulses with finite frequency content. For this, we first compute ensemble correlations $C_\text{pw}(\mathbf{x}_A,\mathbf{x}_B,t; \mathbf{m}^\text{init})$ for a plausible initial model vector $\mathbf{m}^\text{init}$, which contains the coefficients of a suitably discretized initial slowness distribution, $c^{-1}(\mathbf{x})$. While this may be computationally expensive, it only has to be done once, because $C_\text{pw}(\mathbf{x}_A,\mathbf{x}_B,t; \mathbf{m}^\text{init})$ can be used as reference for all subsequent tomographic inversions, e.g., on different patients. The arrival time difference of a wave pulse in $C_\text{pw}(\mathbf{x}_A,\mathbf{x}_B,t; \mathbf{m}^\text{init})$ and its corresponding pulse in the observed random wavefield correlations $C_\text{pw}^\text{obs}(\mathbf{x}_A,\mathbf{x}_B,t)$ can be estimated robustly by cross-correlation \citep{VanDecar_1990},
\begin{equation}\label{E:300}
\Delta t_{AB}^\text{obs} = \text{arg max}\, \int C_\text{pw}(\mathbf{x}_A,\mathbf{x}_B,\tau; \mathbf{m}^\text{init})\, C_\text{pw}^\text{obs}(\mathbf{x}_A,\mathbf{x}_B,t+\tau)\, d\tau\,.
\end{equation}
In heterogeneous media, a time window may need to be applied in order to isolate the correct pulse. In contrast to the picking of individual arrival times, the correlation integral (\ref{E:300}) acts to further suppress the influence of incoherent noise. Facilitating tomographic reconstructions, the correlation time shift depends nearly linearly on sound speed variations of up to 10 \% \citep{Mercerat_2013}.

The measurement error in $\Delta t_{AB}^\text{obs}$ has three major contributions: (1) noise caused by the instrument itself and by surrounding acoustic sources, (2) convergence failures of the correlation function towards the inter-station Green's function related to the unavoidably finite number of transducers and random wavefield realizations, and (3) random errors caused by the random selection of source-time functions for $C_\text{pw}(\mathbf{x}_A,\mathbf{x}_B,t; \mathbf{m}^\text{init})$ and $C_\text{pw}^\text{obs}(\mathbf{x}_A,\mathbf{x}_B,t)$. To eliminate the latter source of errors, we choose identical random realizations of source-time functions for both computed and measured random wavefields. Hence, $C_\text{pw}(\mathbf{x}_A,\mathbf{x}_B,t; \mathbf{m}^\text{init})$ and $C_\text{pw}^\text{obs}(\mathbf{x}_A,\mathbf{x}_B,t)$ are directly comparable, regardless of their convergence towards an inter-receiver Green's function.

\section{Travel time benchmarks}\label{sec:Benchmarks}

The usefulness of the approach proposed in the previous sections critically relies on the accuracy of travel time difference measurements. Hence, before embarking on a study of tomographic reconstructions in section \ref{sec:Tomography}, we analyze travel time differences for a range of different setups. Though such an analysis cannot be exhaustive, it should still point towards useful sets of acquisition parameters.

\subsection{Semi-analytical benchmarks}\label{sec:benchmark}

To assess the performance of the approach proposed in section \ref{sec:theory} under ideal circumstances, we return to the 2-D configuration in Fig.\,\ref{F:summary_correlations} with a homogeneous medium, which allows us to compute Green's functions and random wavefield correlations analytically. Our interest is in the time shift error caused by imperfect convergence towards the Green's function. For this, we compute artificial observations using a homogeneous medium with velocity $c_\text{obs}=1550$ m$/$s. Given the reference medium with $c=1500$ m$/$s and the receiver spacing of $0.15$ m, the expected travel time difference is $0.00322$ ms.

The extent to which this travel time difference can be reproduced by the correlation of random wavefield correlations primarily depends on the number of sources $N_\text{src}$ and the number of realizations $N$. The latter trades off with the length $T$ of an individual realization, which we fix to $0.1$ ms. 

\begin{figure}
\includegraphics[width=\textwidth]{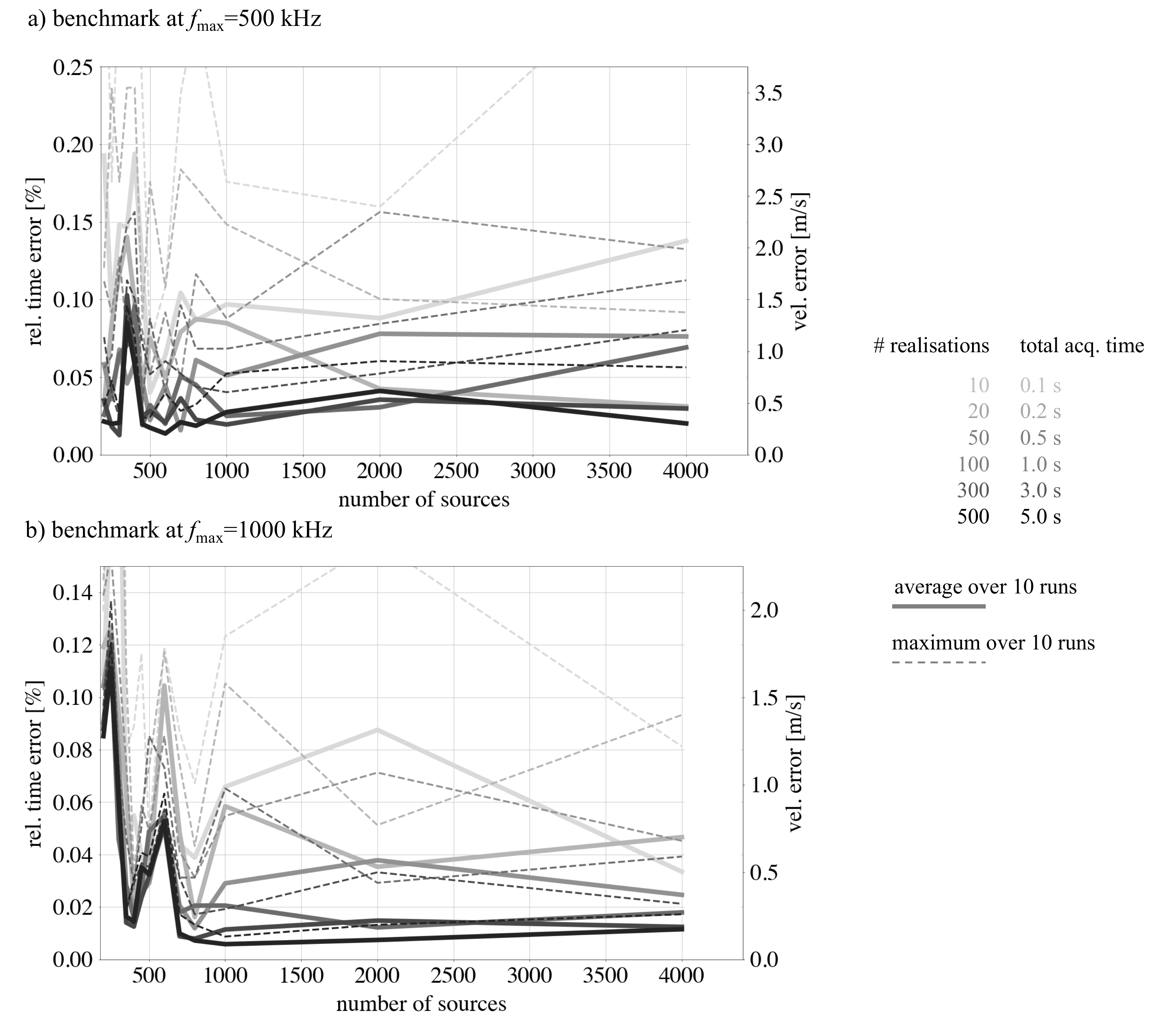}
\caption{Errors in the measured travel time difference relative to the travel time of $0.1$ ms in the reference medium (left axes), and the corresponding velocity errors (right axes). Different gray shading indicates a variable number of realizations $N$, ranging from $10$ (light grey) to $500$ (black). Thick solid curves represent averages over $10$ independent runs, and thin dashed curves mark the maximum error within this ensemble. Panel (a) is for a maximum frequency of 500 kHz, and panel (b) for a maximum frequency of $1$ MHz.[R]}
\label{F:benchmark}
\end{figure}

Fig.\,\ref{F:benchmark} summarizes the travel time errors as a function of $N_\text{src}$ and $N$ for an ensemble of $10$ independent runs. Overall, travel time errors decrease with increasing $N_\text{src}$ and $N$. As expected for a stochastic process, where convergence errors are typically proportional to $1/\sqrt{N_\text{src}}$, rapid initial error reductions are followed by slower improvements that give the appearance of plateauing when plotted on a linear scale. Furthermore, convergence is not uniform, as specific source configurations produce errors that are larger or smaller. This effect results from systematic differences in convergence speed, which depends on a particular distribution of the discrete wavefield sources within the (higher-order) Fresnel zones. A fortunate configuration will lead to an efficient cancellation of sources outside the stationary-phase region (first Fresnel zone), and vice versa.

As order of magnitude, travel time difference errors relative to the exact travel time of $0.1$ s are roughly on the order of $0.05$ \% for a maximum frequency of $500$ kHz. The corresponding velocity errors, averaged over the inter-receiver distance of $0.15$ m are around $1$ m$/$s. These errors can be reduced by a factor of $\sim 2$ when the maximum frequency is $1$ MHz.

\begin{figure}
\includegraphics[width=\textwidth]{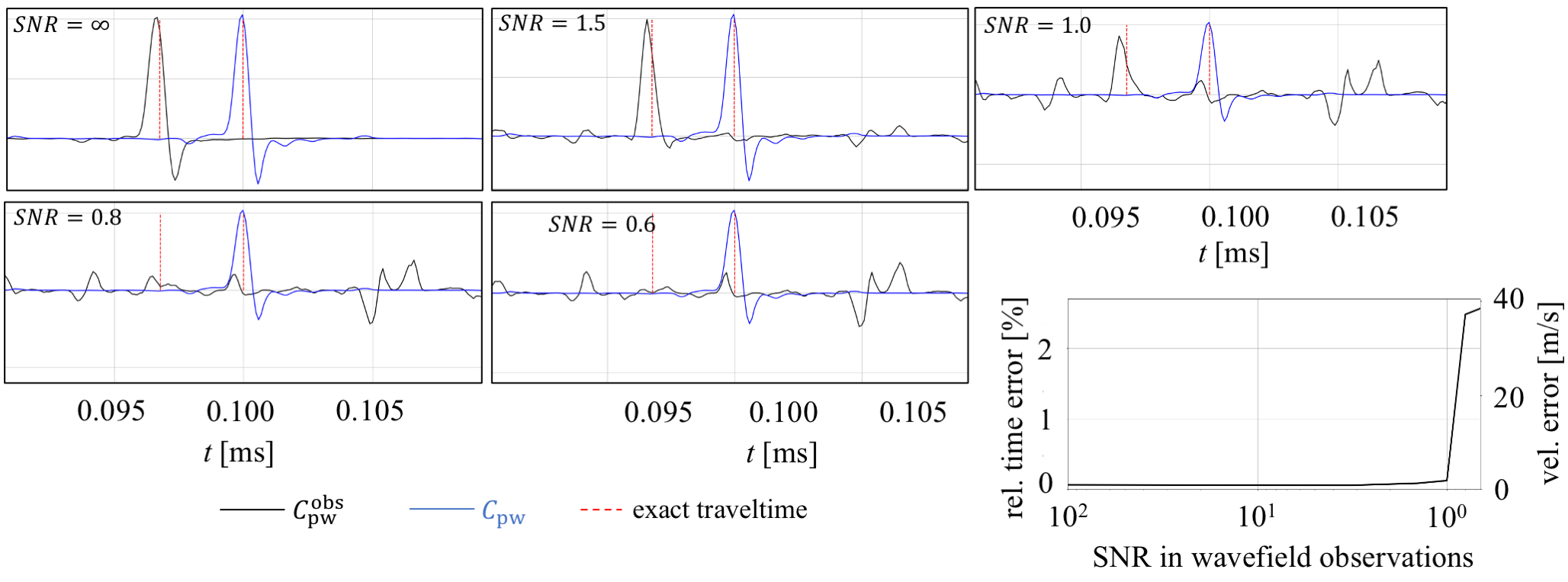}
\caption{Random wavefield correlations for various signal-to-noise ratios (SNR) in the artificial observed wavefields. Artificial observed correlations are shown in blue, their corresponding computed correlations in black. The panel in the lower right summarizes the relative time shift error and the absolute velocity error as a function of the SNR in the random wavefield observations.[R]}
\label{F:snr}
\end{figure}

Correlation and averaging naturally suppress instrumental noise that is non-propagating and incoherent. This is illustrated in Fig.\,\ref{F:snr}, which displays correlation functions for variable signal-to-noise ratios (SNR) in the artificial observed random wavefields for a scenario with $N_\text{src}=500$ and $N=100$. Artificial instrumental noise was mimicked by computing realizations of normally distributed Gaussian noise for each time sample, and then adding a lowpass-filtered version with 1 MHz cutoff to the artificial data. A low SNR of $1.5$ in the wavefield recordings $p(\mathbf{x}_{A,B})$ leads to an SNR of $\sim 20$ in the inter-receiver correlation $C_\text{pw}^\text{obs}(\mathbf{x}_A,\mathbf{x}_B)$ and to a relative error in the measured time shift of merely $0.08\,\%$. The time shift errors only become more significant when the SNR reaches a level where cycle skips occur. In our example, this happens when the SNR in the observed wavefield is around $0.6$.

\subsection{2D numerical breast phantom}\label{sec:2DphantomStudy}

\begin{figure}[!h]
\includegraphics[width=\textwidth]{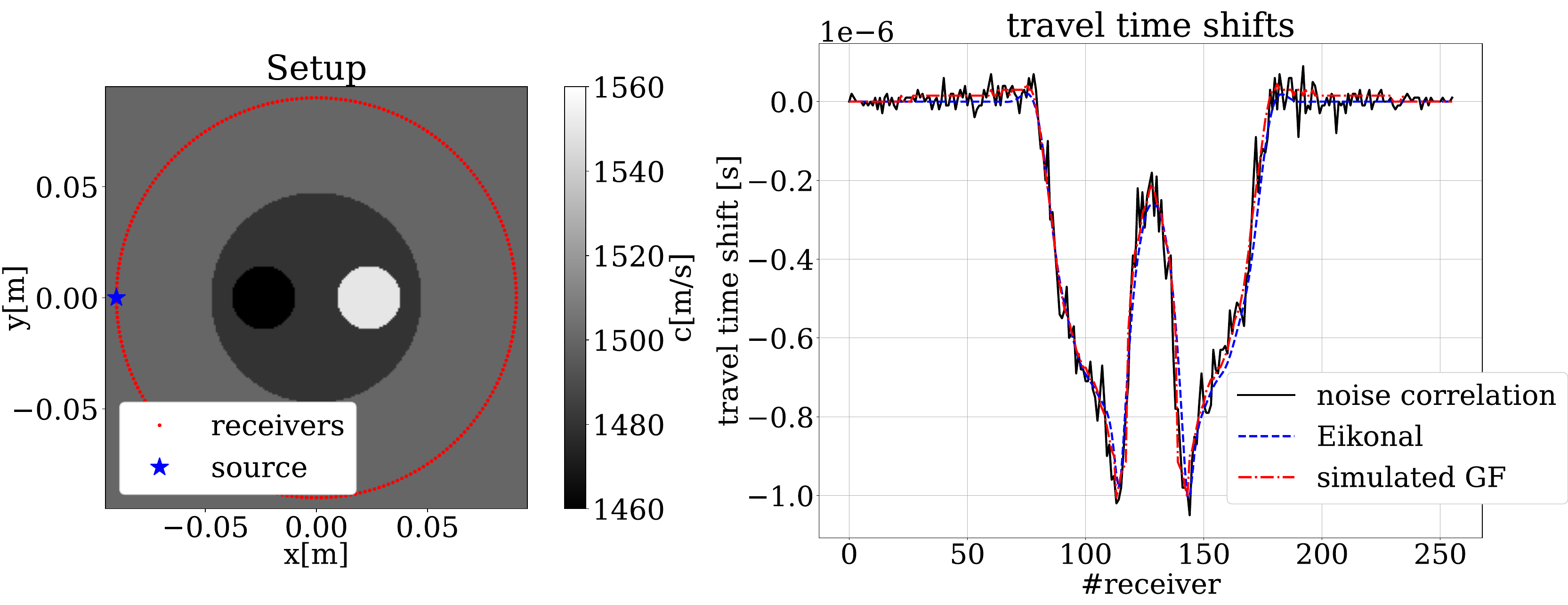}
\caption{(Left) Setup for the numerical simulations. (Right) Comparison of travel time differences obtained from interferometric cross-correlations (black solid curve), bent-ray simulations (blue dashed curve) and simulations of the full wavefield for the direct Green's functions (red curve) between all receiver pairs and a source located at $\star$.} 
\label{F:setupComparisonPlot}
\end{figure}

Taking the semi-analytical implementation described in section \ref{sec:benchmark} one step further, we measure travel time shifts for a 2-D heterogeneous breast phantom, shown in the right panel of Fig.\,\ref{F:setupComparisonPlot}. It consists of a circular area with sound speed $c=$1480 m/s, which is slightly lower than the surrounding values that mimic the water tank. The central part of the phantom contains a low- and and high-speed inclusion. The transducer array with 90 mm radius includes 256 sources that simultaneously act as receivers.

We compute numerical Green's functions between all source-receiver pairs for the heterogeneous phantom using the spectral-element solver Salvus \citep{Salvus} in the frequency range of 75 kHz - 2 MHz and with absorbing boundaries surrounding the domain. To construct the diffuse wavefield, we convolve the filtered Green's functions with a random source-time function, leading to individual realizations of 0.19 ms duration. Subsequently, we compute inter-receiver correlations and phase weights for all receiver pairs. This process repeats $N=500$ times, which translates to a total acquisition time of 0.095 s for the phase-weighted ensemble correlations. 

The spectral-element simulations of ultrasonic wave propagation ensure accurate solutions for complex numerical phantoms. However, the elevated computational cost also limits the number of scenarios that can be considered.

To establish a baseline, we measure travel time differences between the heterogeneous phantom and the homogeneous reference using actual numerical (spectral-element) Green's functions computed with Salvus. Neglecting small numerical errors, the resulting travel time difference curve, shown in Fig.\,\ref{F:setupComparisonPlot}, can serve as ground truth. The corresponding travel time differences obtained through random wavefield interferometry are superimposed in black. They closely follow the baseline, but reveal small fluctuations between neighboring receivers, which we expect from a random process with a finite number of samples. The amplitude of these fluctuations may be reduced by increasing the number of wavefield realizations, and this should be done as a function of the tomographic resolution that one would like to achieve or that is technically achievable with a given number of transducers. As later shown in section \ref{sec:Tomography}, the random fluctuations in this example have significantly less impact than subjective choices of regularization.

To complete this analysis, we complement Fig.\,\ref{F:setupComparisonPlot} with a travel time difference curve from a fast-marching solver of the eikonal equation \citep{Sethian,RickettFomel}. Again, the result closely follows the baseline, as expected. Small systematic differences can, however, be observed at receivers where the corresponding rays travel close to the boundary of a heterogeneity because ray theory ignores the non-zero width of Fresnel zones. 

\section{Tomographic inversions}\label{sec:Tomography}

Eq. \eqref{eq:wave_equation_frequency_domain} provides an accurate description of ultrasonic wave propagation in heterogeneous media, and it may be used directly to solve medical ultrasound full-waveform inversion problems in either the frequency or (after inverse Fourier transform) time domain \citep[e.g.][]{Calderon,Pratt,Perez-Liva,Boehm_2018b}. However, solving the full wave equation repeatedly is computationally challenging, especially in clinical applications. We therefore adopt the ray approximation where space-dependent travel times, $t(\mathbf{x})$, are solutions of the eikonal equation \citep{Cerveny_2001}
\begin{equation}
    |\nabla t(\mathbf{x})^2|=c(\mathbf{x})^{-2}\,.
    \label{eq:Eikonal}
\end{equation}
Eq.\,\eqref{eq:Eikonal} constitutes the forward problem, the solution of which provides calculated travel time differences $\Delta t_{AB}(\mathbf{m})=t_{AB}(\mathbf{m})-t_{AB}^\text{init}(\mathbf{m})$ between discretized versions of a variable slowness distribution $\mathbf{m}$ and the initial slowness distribution $\mathbf{m}^\text{init}$. To estimate an $\mathbf{m}$ that explains observed travel time differences to within their uncertainties, we minimize the misfit functional
\begin{equation}
    J(\mathbf{m})=\frac{1}{2} \sum_{i,j} \frac{1}{\gamma_{ij}^2} [\Delta t_{ij}(\mathbf{m})-\Delta t_{ij}^\text{obs}(\mathbf{m})]^2\,,
    \label{eq:misfit_functional}
\end{equation}
where the sum is over all contributing virtual source-receivers pairs, and $\gamma_{ij}$ denotes the standard deviation of the measurement errors, assumed to follow a normal distribution. 

Numerical solutions of (\ref{eq:Eikonal}) can be computed with a variety of methods with different levels of complexity and accuracy, ranging from straight-ray solvers that ignore ray bending, to fast-marching methods that handle strong heterogeneities and shadow zones \citep{Sethian,RickettFomel}. In the following paragraphs, we will first present a sequence of tomographic reconstructions under the straight-ray approximation. This is intended to highlight the influence of regularization and parametrization. Finally, we employ an eikonal solver that accounts for ray bending, thereby reducing some of the artifacts that result from the straight-ray approximation.

\subsection{Straight-ray tomography}\label{sec:straight}

Under the straight-ray approximation, that is, in a perfectly linear regime, the maximum-likelihood model can be obtained by a simple computation of the generalized inverse \citep[e.g.][]{Fichtner_book_2021}. Since Fig.\,\ref{F:setupComparisonPlot} indicated that the interferometric travel time differences are sufficiently accurate to attempt a tomographic reconstruction, we proceed with computing all  cross-correlation pairs. For this, only one forward simulation with the phantom is required and no calibration data set. Knowing the source and receiver positions, the reference solution for a homogeneous initial model can be computed analytically, and the resulting ensemble correlations can be reused for subsequent tomographic inversions.

\begin{figure}[!h]
\includegraphics[width=1.0\textwidth]{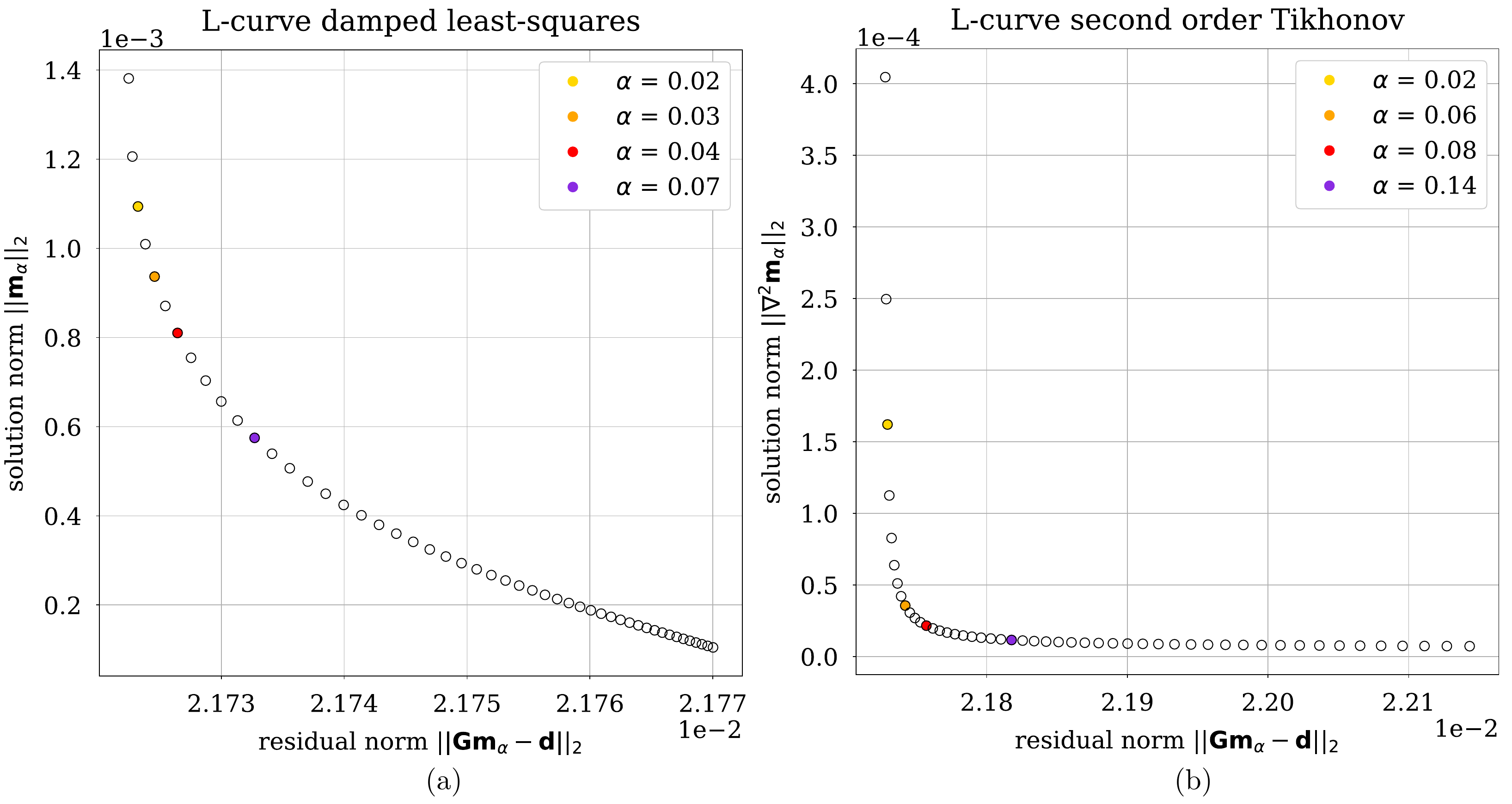}
\caption{L-curve for damped least-squares regularization (a) and Tikhonov regularization (b) using different values of the regularization weight $\alpha$.[R]} 
\label{F:L-curve}
\end{figure}

In a generic setting, the inverse problem defining the tomographic inversions can be written as 
\begin{equation}
    \min_{\mathbf{m}}\frac{1}{2}||\mathbf{G}(\mathbf{m})-\mathbf{d}||_2^2+\alpha\mathbf{R}(\mathbf{m}),
    \label{eq:generic_inverse_prob}
\end{equation}
where $\mathbf{G}$ is the forward operator that maps the observables $\mathbf{d}$, which are the time-of-flight differences, to the a specific model $\mathbf{m}$. Due to the non-unique nature of the inverse problem, some form of regularization is required, which is introduced in equation \eqref{eq:generic_inverse_prob} through the term $\mathbf{R}(\mathbf{m})$ and the regularization weight $\alpha$ that balances the contributions of data misfits and regularization. In the following, we first consider the linearized forward problem $\mathbf{G}\mathbf{m}$ using the straight-ray-tracing forward operator $\mathbf{G}$ and compare inversion results for three different formulations of the regularization term, namely damped least-squares, second-order Tikhonov regularization and total-variation (TV) regularization. For damped least-squares and Tikhonov regularization, the regularization term $\mathbf{R}(\mathbf{m})$ is linear and given explicitly by a matrix, hence a closed-form solution to the inverse problem \,\eqref{eq:generic_inverse_prob} can be formulated as
\begin{equation}
    \mathbf{\tilde{m}}=\bigg(\mathbf{G}(\mathbf{m})^T\mathbf{G}(\mathbf{m})+\alpha\mathbf{R^{\prime}}(\mathbf{m})^T\mathbf{R^{\prime}}(\mathbf{m})\bigg)^{-1}\mathbf{G}(\mathbf{m})^T\mathbf{d},
    \label{eq:regularized_LS_solution}
\end{equation}
where $\mathbf{R^{\prime}}(\mathbf{m})$ is the derivative of the regularization term with respect to $\mathbf{m}$ defined as $R^{\prime}_{ij}=\frac{\partial R_i}{\partial m_j}$ and $\mathbf{\tilde{m}}$ denotes the optimal model in a least-squares sense. Useful levels of regularization can be obtained by examining the trade-off between the solution norm and the residual norm as a function of the regularization weight $\alpha$ as shown in Fig.\,\ref{F:L-curve}. Acceptable regularization weights are typically found around the kink of the L-curve. Below, we compare the result for four different regularization weightings, indicated in Fig.\,\ref{F:L-curve} by the colored dots. 

\begin{figure}[!h]
\centering
\includegraphics[width=0.7\textwidth]{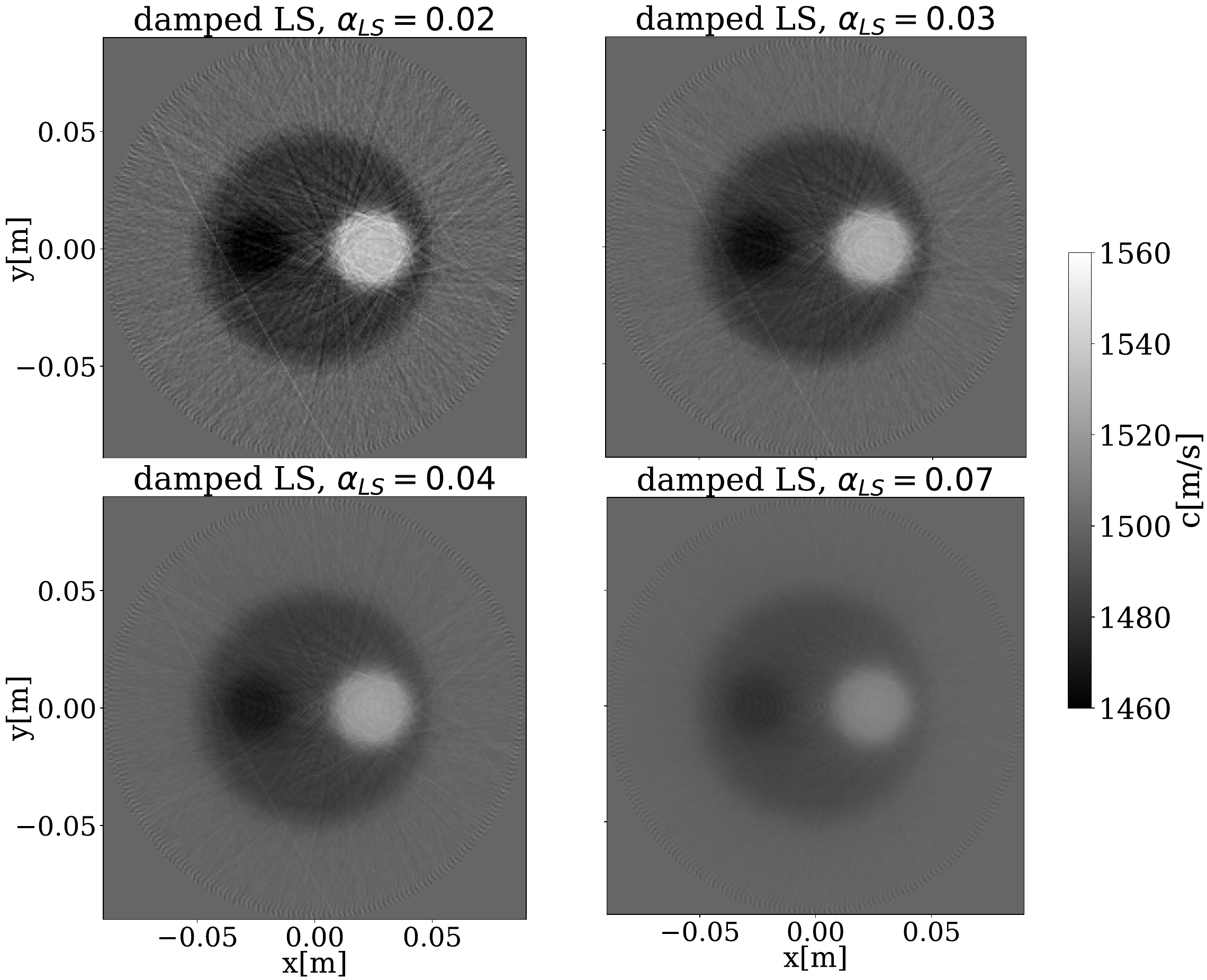}
\caption{Reconstructed sound speed maps using damped least-squares.[R]}\label{F:dampedLS}
\end{figure}

\begin{figure}[!h]
\centering
\includegraphics[width=0.7\textwidth]{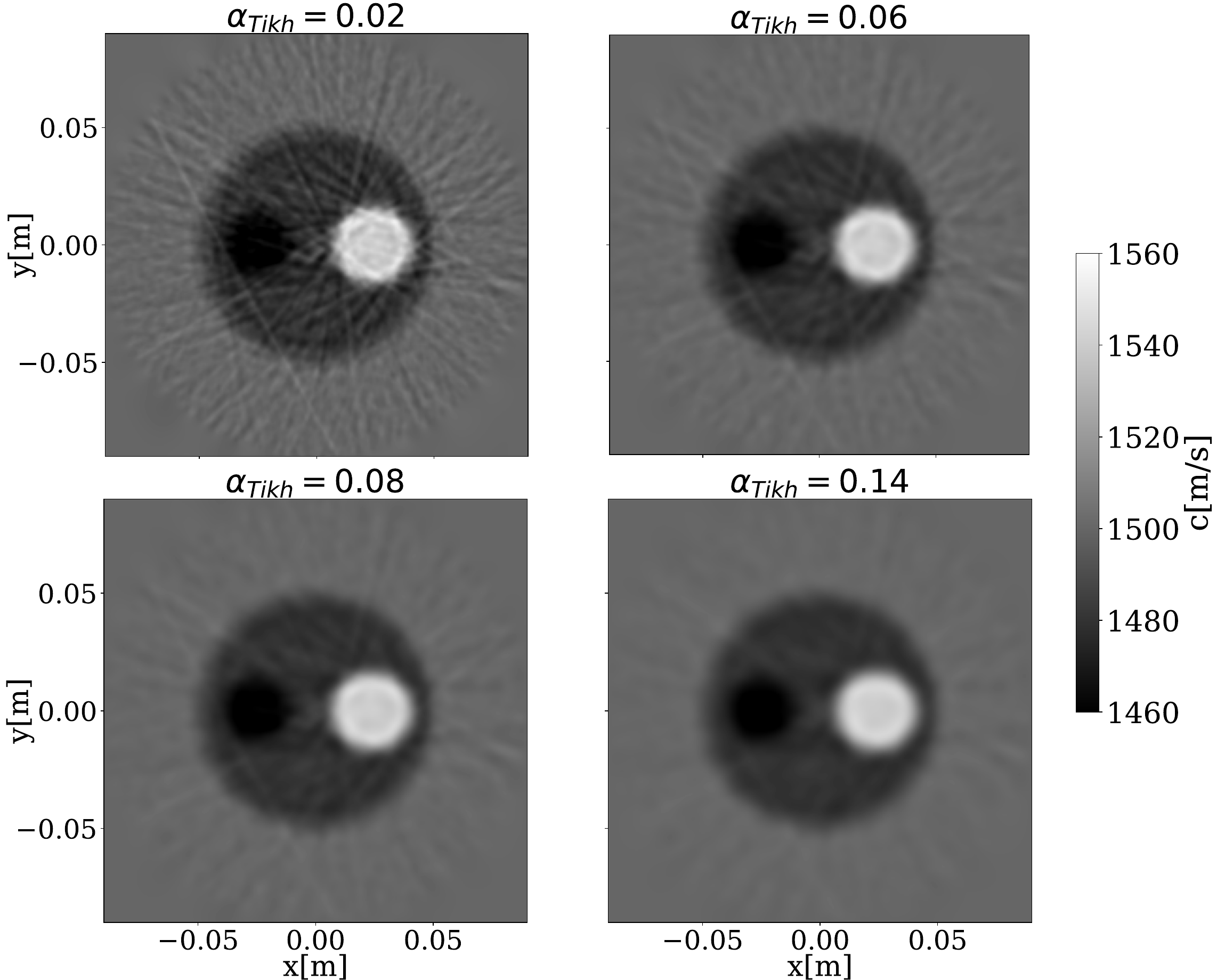}
\caption{Reconstructed sound speed maps using second-order Tikhonov regularization for different regularization weightings $\alpha_{\text{Tikh}}$.[R]} 
\label{F:Tikh}
\end{figure}

The simple damped least-squares formulation applies a damping weight $\alpha$ directly to the model norm, hence $\mathbf{R}(\mathbf{m})=\frac{1}{2}||\mathbf{m}||_2^2$ in eq.\,\eqref{eq:generic_inverse_prob}. The resulting reconstruction in Fig.\,\ref{F:dampedLS} already distinguishes the sound speed variations in the numerical phantom. However, the damping penalises deviations from the background medium. This is particularly visible in the lower right panel of Fig.\,\ref{F:dampedLS}, where the strong damping limits the deviation of the sound speed values from the background medium to a small range. 

Second-order Tikhonov (smoothing) and total-variation regularization largely circumvent this well-known problem of least-squares damping \citep[e.g.][]{Vogel}. Using second-order Tikhonov, the regularization term applies the $\ell_2$ norm to the spatial Hessian with respect to the model parameters, thus, $\mathbf{R}(\mathbf{m})=\frac{1}{2}||\nabla^2\mathbf{m}||^2_2$. The smoothing effect of the spatial Hessian is clearly visible in Fig.\,\ref{F:Tikh}, where ray artifacts are successfully eliminated by stronger regularization weights. The latter have been chosen on the basis of the L-curve plots in Fig.\,\ref{F:L-curve} (b).

\begin{figure}[!h]
\centering
\includegraphics[width=0.8\textwidth]{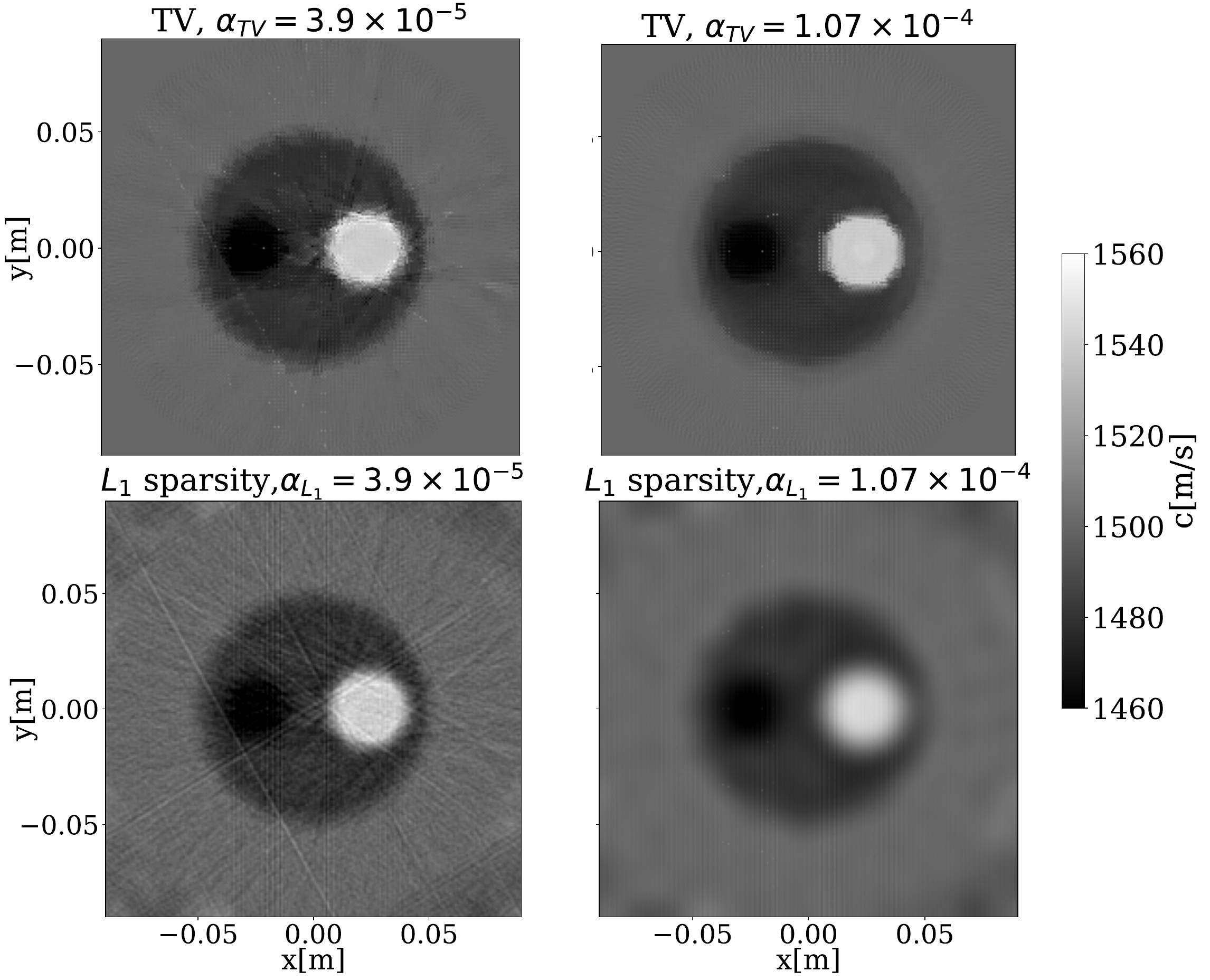}
\caption{Reconstructed speed-of-sound maps with TV regularization in pixel space (top row) and a regularization term equal to the $\ell_1$ norm applied to the sparse fourier transformed model vector (bottom row) for different values of the regularization weight $\alpha$.[R]} 
\label{F:TVsparsity}
\end{figure}

By design, total-variation regularization successfully reconstructs the discontinuous sound speed distribution of the phantom by penalizing the $\ell_1$ norm of the spatial gradient, hence, $\mathbf{R}(\mathbf{m})=||\nabla\mathbf{m}||_1$. In the discrete setting, the gradient may be approximated by a first-order finite-difference matrix. For models structured in locally homogeneous regions with discontinuities on their boundaries, the application of a first-order finite-difference matrix to $\mathbf{m}$ with subsequent summation of the absolute value over all pixels in the grid will result in a gradient only sparsely populated with non-zero values. Hence jumps in the gradient characterizing sharp edges are promoted, which explains the clear delineation of the two inclusions in Fig.\,\ref{F:TVsparsity}. To deal with the non-smooth $\ell_1$ norm in the regularization term of the total variation formulation, we use the Split-Bregman method \citep[][]{PyLops}, which considers the objective
\begin{equation}
    \min_{\mathbf{m}}\frac{1}{2}||\mathbf{G}(\mathbf{m})-\mathbf{d}||_2^2+\alpha||\nabla\mathbf{m}||_1.
    \label{eq:TV_SplitBregman}
\end{equation}
The influence of applying a stronger regularization weight $\alpha$ is especially visible in the upper right plot in Fig.\,\ref{F:TVsparsity}, where the piece-wise nature of the phantom is particularly well reconstructed. 


The bottom row of Fig. \ref{F:TVsparsity} also shows the influence of exploiting the sparsity property of the Fourier basis in the model space. To promote sparse solutions with only a small number of non-zero coefficients in the Fourier space, the transform $\mathbf{P}\mathbf{m}$, where $\mathbf{P}$ is the Fourier operator, enters as $\ell_1$ penalty into the objective such that the second term in eq.\,\eqref{eq:TV_SplitBregman} reads $\mathbf{R}(\mathbf{m})=||\mathbf{P}\mathbf{m}||_1$ \citep[][]{Ulrich}. Transforming the problem to a domain where the structure of the medium can be represented by a sparse set of basis functions generally compresses information to a much smaller set of coefficients. However, the lower right plot in Fig. \ref{F:TVsparsity}, where the regularization contribution is increased in comparison to the lower left plot, shows the stronger influence of the Fourier basis, which is by construction smooth.

\subsection{Bent-ray tomography}\label{sec:bent}

In the straight-ray reconstructions, the heterogeneity with high sound speed consistently appears smaller than its neighbor with lower sound speed, even though they have equal size in the phantom. The removal of this artifact requires the incorporation of ray bending, which introduces non-linearity into the inverse problem. Hence, the minimization of the misfit functional $J$ proceeds iteratively, using gradients computed via adjoint techniques \citep[e.g.][]{SethianPopovici} applied to a fast-marching eikonal solver \citep[e.g.][]{SethianPopovici}. To iteratively update $\mathbf{m}$, starting from $\mathbf{m}^\text{init}$, we use the L-BFGS algorithm \citep[e.g.][]{NocedalWright}
\begin{equation}
    \mathbf{m}_{k+1}=\mathbf{m}_k-\alpha_k\mathbf{H}_k^{-1}\nabla J(\mathbf{m}_k)\,,
    \label{eq:BFGS_update}
\end{equation}
where $\mathbf{m}_k$ is the slowness model in the $k^\text{th}$ iteration, $\alpha_k$ is a step length satisfying the Wolfe condition, $\mathbf{H}_k^{-1}$ is an approximate inverse Hessian of $J(\mathbf{m}_k)$, and $\nabla J(\mathbf{m}_k)$ is the gradient of the misfit functional evaluated at the current model. During each iteration, the approximate inverse Hessian is updated, incorporating new information gained by the model update. Eq.\eqref{eq:misfit_functional} together with eq.\,\eqref{eq:BFGS_update} describe how travel time difference measurements translate into a velocity model that eventually explains observed data to within the observational errors. 

\begin{figure}[!h]
\centering
\includegraphics[width=0.7\textwidth]{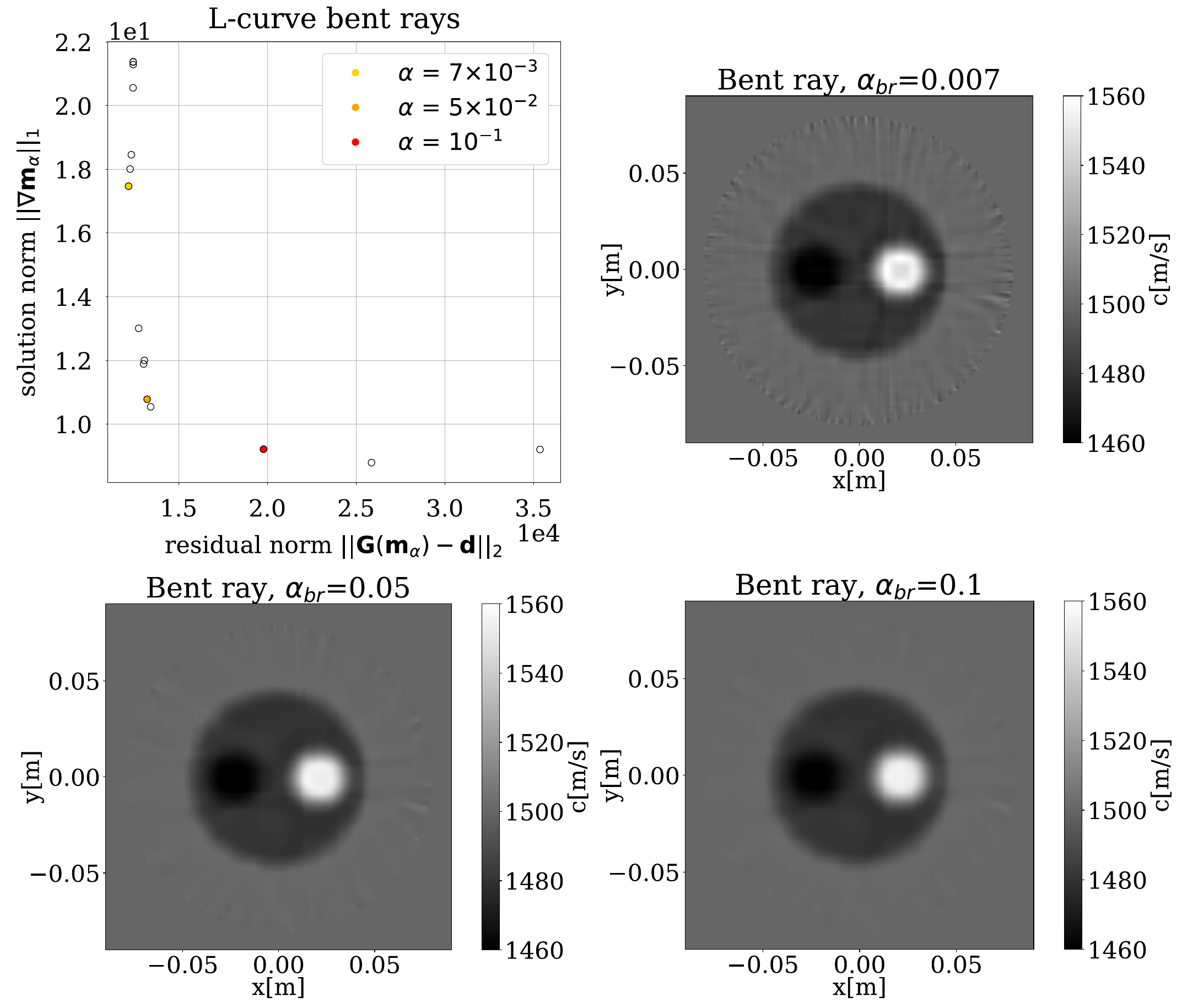}
\caption{Reconstructed sound speed maps using a bent-ray algorithm that solves the eikonal equation and uses total-variation regularization with three different regularization weights $\alpha_{br}$}. 
\label{F:bentray}
\end{figure}

Fig.\,\ref{F:bentray} displays reconstructed sound speed maps using a bent-ray forward model and total-variation regularization after 40 iterations of the L-BFGS algorithm and for different regularization weights. In contrast to the straight-ray reconstructions in Figs. \ref{F:dampedLS}, \ref{F:Tikh} and \ref{F:TVsparsity}, the bent-ray forward model successfully reconstructs the sizes of the two inclusions.

\section{Discussion}\label{sec:Discussion}

We presented an alternative conceptual approach to USCT that aims to reduce acquisition time and eliminate the calibration step. The main methodological ingredients are the following: (1) Active generation of a diffuse acoustic wavefield by simultaneously acting sources. (2) Computation of ensemble-averaged inter-receiver correlations using phase-weighted stacking. (3) Numerical calculation of reference correlations for an initial sound speed distribution and identical random source realizations as for the real-data acquisition. (4) Estimation of travel time differences by cross-correlation and their (iterative) inversion for the sound speed distribution.

In the next paragraphs, we discuss the niche where the proposed method may be beneficial, as well as key advantages and limitations.

\subsection{Potential savings in acquisition time and scaling of the problem}

The relative efficiency of the proposed method primarily depends on its scaling properties, the noise characteristics of the acquisition system, and the desired tomographic resolution. The total acquisition time $\bar{T}_\text{rand}$ is
\begin{equation}\label{E:discussion001}
\bar{T}_\text{rand}=N_\text{rand} T_\text{rand}\,,
\end{equation}
with the number of random wavefield realizations $N_\text{rand}$ and the duration of an individual realization $T_\text{rand}$. By design, $\bar{T}_\text{rand}$ is independent of the number of sources $N_\text{src}$, in contrast to conventional deterministic approaches where sources act sequentially, leading to a total acquisition time of
\begin{equation}\label{E:discussion002}
\bar{T}_\text{det} = N_\text{src} N_\text{det} T_\text{det}\,.
\end{equation}
In (\ref{E:discussion002}), $T_\text{det}$ denotes the time between subsequent shots, and $N_\text{det}$ is the number of repetitions, needed to achieve a desired signal-to-noise ratio through stacking. Since the effects of changing $T_\text{rand}$ and $N_\text{rand}$ trade off exactly in the random wavefield approach, we can assume $T_\text{rand}=T_\text{det}$, without loss of generality, which leads to
\begin{equation}\label{E:discussion003}
\frac{\bar{T}_\text{det}}{\bar{T}_\text{rand}} = N_\text{src} \frac{N_\text{det}}{N_\text{rand}}\,.
\end{equation}
Eq.\,(\ref{E:discussion003}) states that the random wavefield approach  outperforms the sequential deterministic approach when the number of sources is large enough. The precise number of sources where $\bar{T}_\text{det}/\bar{T}_\text{rand}=1$ is controlled by the ratio $N_\text{det}/N_\text{rand}$, which, in turn, depends on the noise characteristics and the desired signal-to-noise ratio. 

It follows that an exact assessment of relative efficiency can only be done on an application-specific basis that accounts for the specifics of an actual acquisition system. However, realizing that tomographic resolution ultimately depends on the number of transducers, the random wavefield approach seems to have considerable potential, as we strive to constrain increasingly small sound speed details.

\subsection{Signal-to-noise ratio}

As stressed in the previous section, the signal-to-noise ratio plays a fundamental role for data quality and scaling properties. Common methods to increase the signal-to-noise ratio include, for instance, averaging over several A-scans or matched filtering, and they may differ for different acquisition systems \citep[e.g.][]{BirkDappRuiter}. 

Conveniently, the ensemble average correlation, as defined in section \ref{SS:practical_calculation_of_xcorr}, already includes an averaging process, which acts to suppress incoherent instrumental noise. The type of 'noise' that is more important for the method presented here is related to insufficient convergence of a stochastic process and the resulting random fluctuations of travel time difference measurements.

\subsection{Elimination of the calibration step}

The diffuse wavefield approach transforms each receiver into a virtual source with precisely known properties. It follows that there is no need for calibration runs in a homogeneous medium (typically water) because there is inherently no need to account for system-specific influences such as the angular dependence of the transducers or system delays \citep{Ruiter}. The latter are, in fact, automatically eliminated by the correlation procedure.

The role of a reference is instead being played by the synthetic random wavefield correlations, computed with exact copies of the actual random source realizations for a numerical reference or initial sound speed model. Since reference correlations can be pre-computed once and for all, travel time differences may be inferred directly for any number of patients screened. In a highly optimized clinical routine, which aims at maximum patient throughput, random wavefield interferometry may therefore offer an essential speed-up compared to standard sequential data acquisition.

\subsection{Finite-frequency travel times}

The measurement process described in section \ref{SS:measurement} defines an arrival time difference between observed and computed wave pulses with finite frequency content. This is not necessarily identical to an arrival time difference in ray theory, upon which we base our tomographic reconstructions. In fact, ray theory assumes infinite frequencies, which do, however, not exist in numerical simulations or actual experiments.

This discrepancy constitutes a systematic error. It vanishes in the hypothetical case where observed and computed wave shapes are identical, that is, for instance, when the actual and the model medium are homogeneous. Generally avoiding this error, requires the replacement of ray theory by finite-frequency theory, which substitutes infinitely thin rays by volumetrically extended sensitivity kernels that correspond exactly to a certain type of finite-frequency measurement, such as travel time differences by cross-correlation \citep[][]{Luo_Schuster_1991,Dahlen_2000,Korta_2020}.

Our acceptance of this systematic error is a pragmatic choice. It is justified by the weak heterogeneities within soft tissue, which, being on the order of few percent, generally do not lead to significant distortions of the transmitted wave pulses \citep{Mercerat_2013}. Furthermore, the use of finite-frequency theory would increase the computational cost of a tomographic reconstruction by orders of magnitude.

\section{Outlook}

This work consitutes a theoretical, and unavoidably simplified, proof of concept of a new USCT approach that we consider a necessary prelude to real-world implementations. Since we consider our results encouraging, the long-term outlook is self-evident. In addition to this, there are variations and adaptations of the method that potentially merit further investigation.

In deterministic sequential acquisition, data coverage is not only controlled by the number of transducers but also by their opening angle. The diffuse wavefield approach, in contrast, may circumvent this limitation because multiple-scattering at the boundaries of the device will eventually produce a nearly equipartitioned wavefield that illuminates the complete medium, independent of the opening angle. The resulting virtual sources will have a full opening angle, thereby increasing coverage and tomographic resolution. A rigorous test of this effect is beyond the scope of this work, as it would require the consideration of a range of device geometries in order to produce meaningful results.

The approach of computing synthetic (reference) and observed random wavefields with exact copies of random source realizations carries the potential to go beyond travel time tomography. In fact, the resulting correlation wavefields are directly comparable in their entirety, which should enable full-waveform inversion methods, similar to seismology \citep{Fichtner_2016,Sager_2018b,Sager_2020}.

\section*{Acknowledgments}
The authors would like to thank Guust Nolet for fruitful discussions about finite-frequency travel time measurements using cross-correlation. This work was supported by the Swiss National Supercomputing Centre (CSCS) under project ID s1040.



\section*{Appendix}
\subsection*{Modal expansion of the Green's function}
\label{sec:appendix}
We consider the acoustic wave equation in the frequency domain as introduced in eq.\,\eqref{eq:wave_equation_frequency_domain}
\begin{equation}
 \frac{\omega^2}{\rho(\mathbf{x}) c^2(\mathbf{x})}p(\mathbf{x},\omega)+\nabla\cdot\bigg(\frac{1}{\rho(\mathbf{x})}\nabla p(\mathbf{x},\omega)\bigg)=-\frac{1}{\rho(\mathbf{x})}f(\mathbf{x},\omega),
 \label{eq:wave_equation_frequency_domain_app}
\end{equation}
where $p(\mathbf{x},\omega)=\int_{-\infty}^{\infty}p(\mathbf{x},t)e^{-i\omega t}dt$. Depending on the specifics of a particular setup, either Neumann, Dirichlet or absorbing boundary conditions may be enforced along different parts of the domain boundary as 
\begin{align}
    p(\mathbf{x},\omega)&=0,\mathbf{x}\in\partial\Omega_\text{Dirichlet},\\
    \nabla p(\mathbf{x},\omega)\cdot\mathbf{n}(\mathbf{x})&=0,\mathbf{x}\in\partial\Omega_\text{Neumann}.
    \label{eq:BC}
\end{align}
where $\mathbf{n}(\mathbf{x})$ is the outward-pointing unit normal. The pressure wavefield $p(\mathbf{x},\omega)$ can be expanded in the normal modes $\phi_n(\mathbf{x})$ of the wave operator as
\begin{equation}
    p(\mathbf{x},\omega)=\sum_na_n(\omega)\phi_n(\mathbf{x}),
    \label{eq:modal_expansion_p_app}
\end{equation}
where $a_n$ are constants denoting the expansion coefficients, which are physically interpreted as the amplitude. The normal modes of the wave operator form a complete basis of the system that are chosen to satisfy the boundary conditions in eq.\,\eqref{eq:BC} and that are orthonormal under the weighted inner product defined by
\begin{equation}
   \langle\phi_m(\mathbf{x})|\phi_n(\mathbf{x})\rangle=\int_{\Omega}\frac{1}{\rho(\mathbf{x})c^2(\mathbf{x})}\phi_n^{\ast}(\mathbf{x})\phi_m(\mathbf{x})d\mathbf{x}=\delta_{mn}, 
   \label{eq:weighted_inner_product}
\end{equation}
where $^\ast$ means complex conjugation. To find the expansion coefficients $a_n$ in eq.\,\eqref{eq:modal_expansion_p}, we take the weighted inner product of the latter with a mode $\phi_m^{\ast}(\mathbf{x})$ and make use of the spatial orthogonality of the modes, which yields
\begin{equation}
    \int_{\Omega}\frac{1}{\rho(\mathbf{x})c^2(\mathbf{x})}\phi_m^{\ast}(\mathbf{x})\sum_na_n(\omega)\phi_n(\mathbf{x})d\mathbf{x}=\sum_na_n(\omega)\delta_{mn}=a_m(\omega).
    \label{eq:expression_for_expansion_coeffs}
\end{equation}
Taking the scalar product of eq.\,\eqref{eq:wave_equation_frequency_domain} with $\phi_m^{\ast}(\mathbf{x})$ and using the expression of the expansion coefficients in eq.\,\eqref{eq:expression_for_expansion_coeffs}, we get
\begin{equation}
    \omega^2a_m(\omega)+\int_{\Omega}\phi_m^{\ast}(\mathbf{x})\nabla\cdot\bigg(\frac{1}{\rho(\mathbf{x})}\nabla p(\mathbf{x},\omega)\bigg)d\mathbf{x}=-\int_{\Omega}\frac{1}{\rho(\mathbf{x})}\phi_m^{\ast}(\mathbf{x})f(\mathbf{x},\omega)d\mathbf{x}.
    \label{eq:scalar_product_acWaveEq_with_phim}
\end{equation}
eq.\,\eqref{eq:scalar_product_acWaveEq_with_phim} is a variational problem, which can be approximated by a finite dimensional problem using the Rayleigh-Ritz method. This allows us to compute the eigenfunctions $\phi_n(\mathbf{x})$ as well as the eigenvalues $\omega^2_n$. Using that the normal mode $\phi_n(\mathbf{x})$ is a solution to the source-free form of the wave equation \eqref{eq:wave_equation_frequency_domain}
\begin{equation}
    \frac{\omega_n^2}{\rho(\mathbf{x})c^2(\mathbf{x})}\phi_n(\mathbf{x})+\nabla\cdot\bigg(\frac{1}{\rho(\mathbf{x})}\nabla\phi_n(\mathbf{x})\bigg)=0,
    \label{eq:normal_mode_solution}
\end{equation}
we obtain an expression for the volume integral on the left-hand side of eq.\,\eqref{eq:scalar_product_acWaveEq_with_phim} in terms of the eigenfrequencies $\omega_n^2$ by taking again the scalar product with $\phi_m^{\ast}(\mathbf{x})$
\begin{equation}
    \omega_n^2\delta_{mn}=-\int_{\Omega}\phi_m^{\ast}(\mathbf{x})\nabla\cdot\bigg(\frac{1}{\rho(\mathbf{x})}\nabla\phi_n(\mathbf{x})\bigg)d\mathbf{x}.
    \label{eq:volume_integral_in_terms_of_eigenfrequencies}
\end{equation}
To shift one spatial derivative to $\phi_m^{\ast}(\mathbf{x})$, we use intergation by parts and exploit that the eigenmodes satisfy the boundary conditions in eq.\,\eqref{eq:BC}
\begin{equation}
    \omega_n^2\delta_{mn}=\int_{\Omega}\frac{1}{\rho(\mathbf{x})}\nabla\phi_m^{\ast}(\mathbf{x})\nabla\phi_n(\mathbf{x})d\mathbf{x}.
    \label{eq:after_intergation_by_parts}
\end{equation}
Substituting eq.\,\eqref{eq:after_intergation_by_parts} into eq.\,\eqref{eq:scalar_product_acWaveEq_with_phim}, we find a representation for the expansion coefficients $a_n$ due to an arbitrary source $f(\mathbf{x},\omega)$
\begin{equation}
    a_m(\omega)=-\frac{\int_{\Omega}\frac{1}{\rho(\mathbf{x})}\phi_m^{\ast}(\mathbf{x})f(\mathbf{x},\omega)d\mathbf{x}}{(\omega^2-\omega_m^2)}.
    \label{eq:expansion_coeffs_arbirtary_source_app}
\end{equation}
Note that each coefficient $a_m$ is weighted by the term $(\omega^2-\omega_m^2)^{-1}$, comprising the angular frequency $\omega$ ,resulting from taking the fourier transform in eq.\,\eqref{eq:wave_equation_frequency_domain}, and the eigenfrequency $\omega_m$, belonging to a specific eigenmode $\phi_m(\mathbf{x})$. For mathematical convenience, we assume that all eigenmodes have distinct and non-degenerated eigenfrequencies. Inserting the expression for the expansion coefficients in eq.\,\eqref{eq:modal_expansion_p} defines an analytical expression of the modal representation of a pressure wavefield recorded at a position $\mathbf{x}$, due to an arbitrary source field $f(\mathbf{x},\omega)$: 
\begin{equation}
    p(\mathbf{x},\omega)=-\sum_n\frac{\int_{\Omega}\frac{1}{\rho(\mathbf{x})}\phi_n^{\ast}(\mathbf{x})f(\mathbf{x},\omega)d\mathbf{x}}{(\omega^2-\omega_n^2)}\phi_n(\mathbf{x}).
    \label{eq:modal_expansion_p_arbitrary_source_app}
\end{equation}

\bibliographystyle{unsrtnat}
\bibliography{references}  

\begin{thebibliography}{54}
\providecommand{\natexlab}[1]{#1}
\providecommand{\url}[1]{\texttt{#1}}
\expandafter\ifx\csname urlstyle\endcsname\relax
  \providecommand{\doi}[1]{doi: #1}\else
  \providecommand{\doi}{doi: \begingroup \urlstyle{rm}\Url}\fi

\bibitem[Greenleaf and Johnson(1975)]{GreenleafJohnson1975}
{J. F.} Greenleaf and {S. A.} Johnson.
\newblock Algebraic reconstruction of spatial distributions of refractive index
  and attenuation in tissues from time-of-flight and amplitude profiles.
\newblock pages 109--119, January 1975.
\newblock Proc of Semin on Ultrason Tissue Charact, 1st, NBS ; Conference date:
  28-05-1975 Through 30-05-1975.

\bibitem[{Greenleaf} and {Bahn}(1981)]{GreenleafBahn1981}
J.~F. {Greenleaf} and R.~C. {Bahn}.
\newblock Clinical imaging with transmissive ultrasonic computerized
  tomography.
\newblock \emph{IEEE Transactions on Biomedical Engineering}, BME-28\penalty0
  (2):\penalty0 177--185, 1981.
\newblock \doi{10.1109/TBME.1981.324789}.

\bibitem[Glover(1977)]{Glover1977}
G.H. Glover.
\newblock Computerized time-of-flight ultrasonic tomography for breast
  examination.
\newblock \emph{Ultrasound in Medicine \& Biology}, 3\penalty0 (2):\penalty0
  117--127, 1977.
\newblock ISSN 0301-5629.
\newblock \doi{https://doi.org/10.1016/0301-5629(77)90064-3}.
\newblock URL
  \url{https://www.sciencedirect.com/science/article/pii/0301562977900643}.

\bibitem[Gemmeke et~al.(2017)Gemmeke, Hopp, Zapf, Kaiser, and
  Ruiter]{KIT_3Dscanner}
Hartmut Gemmeke, Torsten Hopp, Michael Zapf, Clemens Kaiser, and Nicole~V.
  Ruiter.
\newblock 3d ultrasound computer tomography: Hardware setup, reconstruction
  methods and first clinical results.
\newblock \emph{Nuclear Instruments and Methods in Physics Research Section A:
  Accelerators, Spectrometers, Detectors and Associated Equipment},
  873:\penalty0 59--65, 2017.
\newblock ISSN 0168-9002.
\newblock \doi{https://doi.org/10.1016/j.nima.2017.07.019}.
\newblock URL
  \url{https://www.sciencedirect.com/science/article/pii/S0168900217307593}.
\newblock Imaging 2016.

\bibitem[Malik et~al.(2018)Malik, Terry, Wiskin, and
  Lenox]{QT_MalikTerryWiskin2018}
Bilal Malik, Robin Terry, James Wiskin, and Mark Lenox.
\newblock Quantitative transmission ultrasound tomography: imaging and
  performance characteristics.
\newblock \emph{Medical Physics}, 45, 05 2018.
\newblock \doi{10.1002/mp.12957}.

\bibitem[Duric et~al.(2013)Duric, Littrup, Schmidt, Li, Roy, Bey-Knight, Janer,
  Kunz, Chen, Goll, Wallen, Zafar, Allada, West, Jovanovic, Li, and
  Greenway]{SoftVue_Duric2013}
Neb Duric, Peter Littrup, Steven Schmidt, Cuiping Li, Olivier Roy, Lisa
  Bey-Knight, Roman Janer, Dave Kunz, Xiaoyang Chen, Jeffrey Goll, Andrea
  Wallen, Fouzaan Zafar, Venkata Satya Veerendra~Prasad Allada, Erik West,
  Ivana Jovanovic, Kuo Li, and William Greenway.
\newblock Breast imaging with the softvue imaging system: First results.
\newblock volume 8675, page 86750K, 03 2013.
\newblock \doi{10.1117/12.2002513}.

\bibitem[Ranger et~al.(2010)Ranger, Littrup, Duric, Li, Schmidt, Lupinacci,
  Myc, Szczepanski, Rama, and Bey-Knight]{SoftVue_MRIcomparison}
Bryan Ranger, Peter Littrup, Neb Duric, Cuiping Li, Steven Schmidt, Jessica
  Lupinacci, Lukasz Myc, Amy Szczepanski, Olsi Rama, and Lisa Bey-Knight.
\newblock {Breast imaging with ultrasound tomography: a comparative study with
  MRI}.
\newblock In Jan D'hooge and Stephen~A. McAleavey, editors, \emph{Medical
  Imaging 2010: Ultrasonic Imaging, Tomography, and Therapy}, volume 7629,
  pages 50 -- 58. International Society for Optics and Photonics, SPIE, 2010.
\newblock \doi{10.1117/12.845650}.
\newblock URL \url{https://doi.org/10.1117/12.845650}.

\bibitem[Ruiter et~al.(2018)Ruiter, Hopp, Zapf, Menshikov, Kaiser, and
  Gemmeke]{KIT_MRIcomparison}
Nicole~V. Ruiter, Torsten Hopp, Michael Zapf, Alexander Menshikov, C.~Kaiser,
  and Hartmut Gemmeke.
\newblock 3d ultrasound computer tomography for breast cancer diagnosis at kit:
  an overview.
\newblock In \emph{Proceedings of the International Workshop on Medical
  Ultrasound Tomography: 1.- 3. Nov. 2017, Speyer, Germany. Hrsg.: T. Hopp},
  pages 205--216. {KIT Scientific Publishing}, 2018.
\newblock ISBN 978-3-7315-0689-8.
\newblock \doi{10.5445/IR/1000079797}.
\newblock 54.02.02; LK 01.

\bibitem[Roy et~al.(2013)Roy, Schmidt, Li, Allada, West, Kunz, and
  Duric]{RoyDuric}
Olivier Roy, Steven Schmidt, Cuiping Li, Venkata Satya Veerendra~Prasad Allada,
  Erik West, David Kunz, and Neb Duric.
\newblock Breast imaging using ultrasound tomography: From clinical
  requirements to system design.
\newblock pages 1174--1177, 07 2013.
\newblock ISBN 978-1-4673-5686-2.
\newblock \doi{10.1109/ULTSYM.2013.0300}.

\bibitem[Claerbout(1968)]{Claerbout}
Jon~F. Claerbout.
\newblock {Synthesis of a layered medium from its acoustic transmission
  response}.
\newblock \emph{Geophysics}, 33\penalty0 (2):\penalty0 264--269, 04 1968.
\newblock ISSN 0016-8033.
\newblock \doi{10.1190/1.1439927}.
\newblock URL \url{https://doi.org/10.1190/1.1439927}.

\bibitem[Lobkis and Weaver(2001)]{LobkinsWeaver2001}
Oleg~I. Lobkis and Richard~L. Weaver.
\newblock On the emergence of the green’s function in the correlations of a
  diffuse field.
\newblock \emph{The Journal of the Acoustical Society of America}, 110\penalty0
  (6):\penalty0 3011--3017, 2001.
\newblock \doi{10.1121/1.1417528}.
\newblock URL \url{https://doi.org/10.1121/1.1417528}.

\bibitem[Wapenaar(2003)]{Wapenaar2003}
Kees Wapenaar.
\newblock Synthesis of an inhomogeneous medium from its acoustic transmission
  response.
\newblock \emph{Geophysics}, 68, 09 2003.
\newblock \doi{10.1190/1.1620649}.

\bibitem[Wapenaar and Fokkema(2006)]{WapenaarFokkema2006}
Kees Wapenaar and Jacob Fokkema.
\newblock Green’s function representations for seismic interferometry.
\newblock \emph{GEOPHYSICS}, 71\penalty0 (4):\penalty0 SI33--SI46, 2006.
\newblock \doi{10.1190/1.2213955}.
\newblock URL \url{https://doi.org/10.1190/1.2213955}.

\bibitem[Malcolm et~al.(2004)Malcolm, Scales, and
  Tiggelen]{MalcolmScalesTiggelen2004}
Alison Malcolm, John Scales, and Bart Tiggelen.
\newblock Extracting the green function from diffuse, equipartitioned waves.
\newblock \emph{Physical review. E, Statistical, nonlinear, and soft matter
  physics}, 70:\penalty0 015601, 02 2004.
\newblock \doi{10.1103/PhysRevE.70.015601}.

\bibitem[Shapiro et~al.(2005)Shapiro, Campillo, Stehly, and
  Ritzwoller]{ShapiroCampillo2005}
Nikolai Shapiro, Michel Campillo, Laurent Stehly, and Michael Ritzwoller.
\newblock High-resolution surface-wave tomography from ambient seismic noise.
\newblock \emph{Science (New York, N.Y.)}, 307:\penalty0 1615--8, 04 2005.
\newblock \doi{10.1126/science.1108339}.

\bibitem[Sabra et~al.(2005)Sabra, Gerstoft, Roux, and Kuperman]{Sabra_2005}
K.~G. Sabra, P.~Gerstoft, P.~Roux, and W.~A. Kuperman.
\newblock {Surface wave tomography from microseisms in Southern California}.
\newblock \emph{Geophys. Res. Lett.}, 32:\penalty0 doi:10.1029/2005GL023155,
  2005.

\bibitem[Stehly et~al.(2009)Stehly, Fry, Campillo, Shapiro, Guilbert, Boschi,
  and Giardini]{Stehly_2009}
L.~Stehly, B.~Fry, M.~Campillo, N.~M. Shapiro, J.~Guilbert, L.~Boschi, and
  D.~Giardini.
\newblock {Tomography of the Alpine region from observations of seismic ambient
  noise}.
\newblock \emph{Geophys. J. Int.}, 178:\penalty0 338--350, 2009.

\bibitem[Saygin and Kennett(2012)]{Saygin_2012}
E.~Saygin and B.~L.~N. Kennett.
\newblock {Crustal structure of Australia from ambient seismic noise
  tomography}.
\newblock \emph{J. Geophys. Res.}, 117:\penalty0 doi:10.1029/2011JB008403,
  2012.

\bibitem[Nakata et~al.(2019)Nakata, Gualtieri, and Fichtner]{Nakata_2019b}
N.~Nakata, L.~Gualtieri, and A.~Fichtner.
\newblock \emph{Seismic Ambient Noise}.
\newblock Cambridge University Press, 2019.

\bibitem[Weaver and Lobkis(2004)]{Weaver_2004}
R.~L. Weaver and O.~I. Lobkis.
\newblock {Diffuse fields in open systems and the emergence of Green's
  function}.
\newblock \emph{J. Acoust. Soc. Am.}, 116:\penalty0 2731--2734, 2004.

\bibitem[Fichtner and Tsai(2019)]{Fichtner_2019b}
A.~Fichtner and V.~Tsai.
\newblock Theoretical foundations of noise interferometry.
\newblock In N.~Nakata, L.~Gualtieri, and A.~Fichtner, editors, \emph{Seismic
  Ambient Noise}, pages 109--143. Cambridge University Press, Cambridge, U.K.,
  2019.

\bibitem[Gilbert(1971)]{Gilbert_normal_modes}
Freeman Gilbert.
\newblock Excitation of the normal modes of the earth by earthquake sources.
\newblock \emph{Geophysical Journal of the Royal Astronomical Society},
  22\penalty0 (2):\penalty0 223--226, 1971.
\newblock \doi{https://doi.org/10.1111/j.1365-246X.1971.tb03593.x}.
\newblock URL
  \url{https://onlinelibrary.wiley.com/doi/abs/10.1111/j.1365-246X.1971.tb03593.x}.

\bibitem[Ardhuin et~al.(2011)Ardhuin, Stutzmann, Schimmel, and
  Mangeney]{Ardhuin_2011}
F.~Ardhuin, E.~Stutzmann, M.~Schimmel, and A.~Mangeney.
\newblock Ocean wave sources of seismic noise.
\newblock \emph{J. Geophys. Res.}, 116:\penalty0 doi:10.1029/2011JC006952,
  2011.

\bibitem[Ermert et~al.(2017)Ermert, Sager, Afanasiev, Boehm, and
  Fichtner]{Ermert_2017}
L.~Ermert, K.~Sager, M.~Afanasiev, C.~Boehm, and A.~Fichtner.
\newblock {Ambient seismic source inversion in a heterogeneous Earth: Theory
  and application to the Earth's hum}.
\newblock \emph{J. Geophys. Res.}, 122:\penalty0 9184--9207, 2017.

\bibitem[Gualtieri et~al.(2019)Gualtieri, Stutzmann, Juretzek, Hadziioannou,
  and Ardhuin]{Gualtieri_2019}
L.~Gualtieri, E.~Stutzmann, C.~Juretzek, C.~Hadziioannou, and F.~Ardhuin.
\newblock {Global-scale analysis and modeling of primary microseisms}.
\newblock \emph{Geophys. J. Int.}, 218:\penalty0 560--572, 2019.

\bibitem[Ardhuin et~al.(2019)Ardhuin, Gualtieri, and Stutzmann]{Ardhuin_2019}
F.~Ardhuin, L.~Gualtieri, and E.~Stutzmann.
\newblock Physics of ambient noise generation by ocean waves.
\newblock In N.~Nakata, L.~Gualtieri, and A.~Fichtner, editors, \emph{Seismic
  Ambient Noise}, pages 109--143. Cambridge University Press, Cambridge, U.K.,
  2019.

\bibitem[Igel et~al.(2021)Igel, Ermert, and Fichtner]{Igel_2021}
J.~Igel, L.~Ermert, and A.~Fichtner.
\newblock {Rapid finite-frequency microseismic noise source inversion at
  regional to global scales}.
\newblock \emph{Geophys. J. Int.}, 227:\penalty0 169--183, 2021.

\bibitem[Schimmel and Paulssen(1997)]{Schimmel_1997}
M.~Schimmel and H.~Paulssen.
\newblock {Noise reduction and detection of weak, coherent signals through
  phase-weighted stacks}.
\newblock \emph{Geophys. J. Int.}, 130:\penalty0 497--505, 1997.

\bibitem[Schimmel et~al.(2011)Schimmel, Stutzmann, and Gallart]{Schimmel_2011}
M.~Schimmel, E.~Stutzmann, and J.~Gallart.
\newblock {Using instantaneous phase coherence for signal extraction from
  ambient noise data at a local to a global scale}.
\newblock \emph{Geophys. J. Int.}, 184:\penalty0 494--506, 2011.

\bibitem[Igel(2016)]{Igel}
Heiner Igel.
\newblock \emph{Computational Seismology: A Practical Introduction}.
\newblock 10 2016.
\newblock ISBN 9780198717409.
\newblock \doi{10.1093/acprof:oso/9780198717409.001.0001}.

\bibitem[VanDecar and Crosson(1990)]{VanDecar_1990}
J.~C. VanDecar and R.~S. Crosson.
\newblock {Determination of teleseismic relative phase arrival times using
  multi-channel cross-correlation and least squares}.
\newblock \emph{Bull. Seis. Soc. Am.}, pages 150--169, 1990.

\bibitem[Mercerat and Nolet(2013)]{Mercerat_2013}
E.~D. Mercerat and G.~Nolet.
\newblock On the linearity of cross-correlation delay times in finite-frequency
  tomography.
\newblock \emph{Geophys. J. Int.}, 192:\penalty0 681--687, 2013.

\bibitem[Afanasiev et~al.(2019)Afanasiev, Boehm, van Driel, Krischer, Rietmann,
  May, Knepley, and Fichtner]{Salvus}
Michael Afanasiev, Christian Boehm, Martin van Driel, Lion Krischer, Max
  Rietmann, Dave~A May, Matthew~G Knepley, and Andreas Fichtner.
\newblock Modular and flexible spectral-element waveform modelling in two and
  three dimensions.
\newblock \emph{Geophysical Journal International}, 216\penalty0 (3):\penalty0
  1675--1692, 2019.
\newblock \doi{10.1093/gji/ggy469}.

\bibitem[Sethian(1996)]{Sethian}
J~A Sethian.
\newblock A fast marching level set method for monotonically advancing fronts.
\newblock \emph{Proceedings of the National Academy of Sciences}, 93\penalty0
  (4):\penalty0 1591--1595, 1996.
\newblock ISSN 0027-8424.
\newblock \doi{10.1073/pnas.93.4.1591}.
\newblock URL \url{https://www.pnas.org/content/93/4/1591}.

\bibitem[Rickett and Fomel(2001)]{RickettFomel}
James Rickett and Sergey Fomel.
\newblock A second-order fast marching eikonal solver.
\newblock 02 2001.

\bibitem[Calderon~Agudo(2017)]{Calderon}
Oscar Calderon~Agudo.
\newblock 3d imaging of the breast using full-waveform inversion.
\newblock 11 2017.

\bibitem[Pratt et~al.(2007)Pratt, Huang, Duric, and Littrup]{Pratt}
Robert Pratt, Lianjie Huang, Neb Duric, and Peter Littrup.
\newblock Sound-speed and attenuation imaging of breast tissue using waveform
  tomography of transmission ultrasound data.
\newblock \emph{Proceedings of SPIE - The International Society for Optical
  Engineering}, 6510, 03 2007.
\newblock \doi{10.1117/12.708789}.

\bibitem[Pérez-Liva et~al.(2017)Pérez-Liva, Herraiz, Udías, Miller, Cox, and
  Treeby]{Perez-Liva}
M.~Pérez-Liva, J.~L. Herraiz, J.~M. Udías, E.~Miller, B.~T. Cox, and B.~E.
  Treeby.
\newblock Time domain reconstruction of sound speed and attenuation in
  ultrasound computed tomography using full wave inversion.
\newblock \emph{The Journal of the Acoustical Society of America}, 141\penalty0
  (3):\penalty0 1595--1604, 2017.
\newblock \doi{10.1121/1.4976688}.
\newblock URL \url{https://doi.org/10.1121/1.4976688}.

\bibitem[Boehm et~al.(2018)Boehm, {Korta-Martiartu}, Vinard, Balic, and
  Fichtner]{Boehm_2018b}
C.~Boehm, N.~{Korta-Martiartu}, N.~Vinard, I.~J. Balic, and A.~Fichtner.
\newblock {Time-domain spectral-element ultrasound waveform tomography using a
  stochastic quasi-Newton method}.
\newblock \emph{SPIE Medical Imaging 2018}, page 92 – 100, 2018.

\bibitem[Cerveny(2001)]{Cerveny_2001}
V.~Cerveny.
\newblock \emph{{Seismic ray theory}}.
\newblock Cambridge University Press, 2001.

\bibitem[Fichtner(2021)]{Fichtner_book_2021}
A.~Fichtner.
\newblock \emph{Lecture Notes on Inverse Theory}.
\newblock Cambridge Open Engage, doi:10.33774/coe-2021-qpq2j, 2021.

\bibitem[Vogel(2002)]{Vogel}
Curtis~R. Vogel.
\newblock \emph{Computational Methods for Inverse Problems}.
\newblock Society for Industrial and Applied Mathematics, 2002.
\newblock \doi{10.1137/1.9780898717570}.
\newblock URL \url{https://epubs.siam.org/doi/abs/10.1137/1.9780898717570}.

\bibitem[Ravasi and Vasconcelos(2020)]{PyLops}
Matteo Ravasi and Ivan Vasconcelos.
\newblock Pylops—a linear-operator python library for scalable algebra and
  optimization.
\newblock \emph{SoftwareX}, 11:\penalty0 100361, 2020.
\newblock ISSN 2352-7110.
\newblock \doi{https://doi.org/10.1016/j.softx.2019.100361}.
\newblock URL
  \url{https://www.sciencedirect.com/science/article/pii/S2352711019301086}.

\bibitem[Ulrich et~al.(2021)Ulrich, Zunino, Boehm, and Fichtner]{Ulrich}
Ines~E. Ulrich, Andrea Zunino, Christian Boehm, and Andreas Fichtner.
\newblock {Sparsifying regularizations for stochastic sample average
  minimization in ultrasound computed tomography}.
\newblock In Brett~C. Byram and Nicole~V. Ruiter, editors, \emph{Medical
  Imaging 2021: Ultrasonic Imaging and Tomography}, volume 11602, pages 194 --
  209. International Society for Optics and Photonics, SPIE, 2021.
\newblock \doi{10.1117/12.2580926}.
\newblock URL \url{https://doi.org/10.1117/12.2580926}.

\bibitem[Sethian and Popovici(1999)]{SethianPopovici}
James Sethian and Alexander Popovici.
\newblock Three dimensional traveltimes computation using the fast marching
  method.
\newblock \emph{Geophysics}, 64:\penalty0 516--523, 03 1999.
\newblock \doi{10.1190/1.1444558}.

\bibitem[Nocedal and Wright(2006)]{NocedalWright}
Jorge Nocedal and {Stephen J.} Wright.
\newblock \emph{Numerical Optimization: Springer Series in Operations Research
  and Financial Engineering}.
\newblock Springer, 2006.
\newblock ISBN 9780387303031.

\bibitem[Birk et~al.(2014)Birk, Dapp, Ruiter, and Becker]{BirkDappRuiter}
Matthias Birk, Robin Dapp, N.V. Ruiter, and J.~Becker.
\newblock Gpu-based iterative transmission reconstruction in 3d ultrasound
  computer tomography.
\newblock \emph{Journal of Parallel and Distributed Computing}, 74\penalty0
  (1):\penalty0 1730--1743, 2014.
\newblock ISSN 0743-7315.
\newblock \doi{https://doi.org/10.1016/j.jpdc.2013.09.007}.
\newblock URL
  \url{https://www.sciencedirect.com/science/article/pii/S0743731513002037}.

\bibitem[Ruiter(2016)]{Ruiter}
Nicole~Valerie Ruiter.
\newblock Dreidimensionale ultraschall-computertomographie: vom konzept zur
  klinischen anwendung, 2016.
\newblock 54.02.02; LK 01.

\bibitem[Luo and Schuster(1991)]{Luo_Schuster_1991}
Y.~Luo and G.~T. Schuster.
\newblock Wave-equation traveltime inversion.
\newblock \emph{Geophysics}, 56:\penalty0 645--653, 1991.

\bibitem[Dahlen et~al.(2000)Dahlen, Hung, and Nolet]{Dahlen_2000}
F.A. Dahlen, S.-H. Hung, and G.~Nolet.
\newblock Fr\'{e}chet kernels for finite-frequency traveltimes -- {I}.
  {T}heory.
\newblock \emph{Geophys. J. Int.}, 141:\penalty0 157--174, 2000.

\bibitem[{Korta Martiartu} et~al.(2020){Korta Martiartu}, Boehm, and
  Fichtner]{Korta_2020}
N.~{Korta Martiartu}, C.~Boehm, and A.~Fichtner.
\newblock {3-D wave-equation-based finite-frequency tomography for ultrasound
  computed tomography}.
\newblock \emph{IEEE Trans. Ultrasonics, Ferroelectrics, and Frequency
  Control}, 67:\penalty0 1332--1343, 2020.

\bibitem[Fichtner et~al.(2017)Fichtner, Stehly, Ermert, and
  Boehm]{Fichtner_2016}
A.~Fichtner, L.~Stehly, L.~Ermert, and C.~Boehm.
\newblock {Generalised interferometry - I. Theory for inter-station
  correlations}.
\newblock \emph{Geophys. J. Int.}, 208:\penalty0 603--638, 2017.

\bibitem[Sager et~al.(2018)Sager, Ermert, Boehm, and Fichtner]{Sager_2018b}
K.~Sager, L.~Ermert, C.~Boehm, and A.~Fichtner.
\newblock Towards full waveform ambient noise inversion.
\newblock \emph{Geophys. J. Int.}, 212:\penalty0 566--590, 2018.

\bibitem[Sager et~al.(2020)Sager, Boehm, Ermert, Krischer, and
  Fichtner]{Sager_2020}
K.~Sager, C.~Boehm, L.~Ermert, L.~Krischer, and A.~Fichtner.
\newblock Global-scale full-waveform ambient noise inversion.
\newblock \emph{J. Geophys. Res.}, 125, 2020.
\newblock \doi{10.1029/2019JB018644}.

\end{thebibliography}

\end{document}